\documentclass[smallcondensed]{svjour3}

\usepackage{amsmath}
\usepackage{url}
\usepackage{bbm}
\usepackage{amssymb}
\usepackage[algoruled,vlined,english,linesnumbered]{algorithm2e}
\usepackage{xspace}
\usepackage{xcolor}
\usepackage{stmaryrd}

\usepackage{graphicx}
\usepackage{wrapfig}
\usepackage{enumerate}
\usepackage{subfigure}
\usepackage{epsfig}

\usepackage{tikz}

\smartqed
\usepackage{mathptmx}

\usepackage[pdftex,%
	pdfauthor={Chistikov, Dimitrova, Majumdar},%
	pdftitle={Approximate Counting in SMT and Value Estimation for Probabilistic Programs}]{hyperref}

\spnewtheorem{prop}{Proposition}{\bfseries}{\itshape}
\spnewtheorem*{remark*}{Remark}{\itshape}{\upshape}

\newcounter{ex}
\stepcounter{ex}

\makeatletter
\begingroup \catcode `|=0 \catcode `[= 1
\catcode`]=2 \catcode `\{=12 \catcode `\}=12
\catcode`\\=12 |gdef|@xcomment#1\end{comment}[|end[comment]]
|endgroup
\def\@comment{\let\do\@makeother \dospecials\catcode`\^^M=10\def\par{}}
\def\begincomment{\@comment\@xcomment}
\makeatother

\newcommand{\newextmathcommand}[2]{%
    \newcommand{#1}{\ensuremath{#2}\xspace}
}

\newcommand{\renewextmathcommand}[2]{%
    \renewcommand{#1}{\ensuremath{#2}\xspace}
}

\newcommand{\cclass}[1]{\mathbf{#1}}
\newcommand{\newcclass}[2]{\newextmathcommand{#1}{\cclass{#2}}}
\newcommand{\renewcclass}[2]{\renewextmathcommand{#1}{\cclass{#2}}}

\renewcclass{\P}{P}
\newcclass{\NP}{NP}
\newcclass{\BPP}{BPP}
\newcclass{\PSPACE}{PSPACE}
\newcclass{\SharpP}{\#P}

\newcommand{\df}{\textit}

\def\set#1{\{#1 \}}
\def\tuple#1{\langle #1 \rangle }
\def\dbrkts#1{\langle \![ #1 ]\!\rangle}

\newcommand{\cproblem}[1]{\text{\textsc{#1}}}
\newcommand{\newcproblem}[2]{\newextmathcommand{#1}{\cproblem{#2}}}

\newcproblem{\SAT}{SAT}
\newcproblem{\SharpSAT}{\#SAT}
\newcproblem{\SharpSMT}{\#SMT}

\newcommand{\sset}{\subseteq}
\newcommand{\true}{\mathit{true}}

\newcommand{\reals}{\mathbb{R}}

\newcommand{\prb}{\mathsf{Pr}}
\renewcommand{\Pr}{\prb}

\newextmathcommand{\program}{\mathcal P}
\newextmathcommand{\Var}{\mathcal X}
\newextmathcommand{\vars}{\mathcal X}
\newextmathcommand{\bfx}{\mathbf x}
\newextmathcommand{\bfz}{\mathbf z}
\newextmathcommand{\1}{\mathbbm 1}
\newextmathcommand{\R}{\mathbb R}
\newextmathcommand{\N}{\mathbb N}
\newextmathcommand{\Q}{\mathbb Q}
\newextmathcommand{\Z}{\mathbb Z}
\newextmathcommand{\B}{\mathbb B}
\newextmathcommand{\Bin}{\{0, 1\}}

\renewextmathcommand{\phi}{\varphi}
\renewcommand{\ge}{\geq}
\renewcommand{\le}{\leq}

\newcommand{\MyPhi}{\mathrm\Phi}
\newcommand{\MyPsi}{\mathrm\Psi}
\newcommand{\MySigma}{\mathrm\Sigma}
\newcommand{\MyOmega}{\mathrm\Omega}

\newextmathcommand{\vinit}{{\mathsf{init}}}
\newextmathcommand{\vacc}{{\mathsf{acc}}}
\newextmathcommand{\vrej}{{\mathsf{rej}}}
\newcommand{\vaccept}{\vacc}
\newcommand{\vreject}{\vrej}

\newextmathcommand{\SKIP}{\mathsf{skip}}
\newextmathcommand{\ASSUME}{\mathsf{assume}}

\newextmathcommand{\ACCEPT}{\mathsf{accept}}
\newextmathcommand{\REJECT}{\mathsf{reject}}

\newextmathcommand{\Accept}{\mathsf{Accept}}
\newextmathcommand{\Term}{\mathsf{Term}}

\newcommand{\Uniform}{\mathsf{Uniform}}

\newextmathcommand{\Theory}{{\mathcal T}}
\newextmathcommand{\F}{{\mathcal F}}
\newcommand{\satisfies}{\models}
\newcommand{\val}[1]{\mathsf{val}(#1)}
\newcommand{\uval}[1]{\overline{\mathsf{val}}(#1)}
\newcommand{\lval}[1]{\underline{\mathsf{val}}(#1)}
\newcommand{\allmod}[1]{\llbracket{#1}\rrbracket}
\newcommand{\mc}[1]{\mathsf{mc}({#1})}
\newcommand{\vc}[1]{\mathsf{vc}({#1})}

\newcommand{\dist}[1]{\mathsf{dist}({#1})}

\newcommand{\chck}{\text{\sc SMT}}
\newcommand{\pickh}{\text{\sc pick-hash}}

\newextmathcommand{\A}{\mathcal A}
\newextmathcommand{\C}{\mathcal C}
\newextmathcommand{\UB}{\mathcal U}
\newextmathcommand{\Estimate}{\mathcal E}
\newextmathcommand{\Hash}{\mathcal H}
\newextmathcommand{\IA}{\mathsf{IA}}
\newextmathcommand{\RA}{\mathsf{RA}}
\newextmathcommand{\IARA}{\mathsf{IA+RA}}
\newcommand{\qge}{\ge^?}
\newextmathcommand{\eps}{\varepsilon}
\renewcommand{\epsilon}{\varepsilon}

\newcommand{\Compute}{\vc}
\newextmathcommand{\exbool}{B}
\newextmathcommand{\exdet}{N}
\newextmathcommand{\exprob}{P}
\newextmathcommand{\phiacc}{\phi_{\mathsf{acc}}}
\newextmathcommand{\phiterm}{\phi_{\mathsf{term}}}
\newextmathcommand{\exaccept}{\mathsf{accepted}}
\newextmathcommand{\exterm}{\mathsf{terminated}}
\newextmathcommand{\extrue}{\mathsf{true}}
\newextmathcommand{\exmultc}{{\mathbf c}}
\newextmathcommand{\acc}{\mathsf{acc}}
\newextmathcommand{\term}{\mathsf{term}}

\newcommand{\dual}[1]{{{#1}^*}}

\newcommand*{\intextheading}[1]{\textbf{#1}}

\newextmathcommand{\True}{\mathsf{true}}

\newenvironment{compactenumerate}{\begin{enumerate}}{\end{enumerate}}

\newextmathcommand{\zerothsize}{\xi}
\newcommand{\Prob}{\Pr}
\newcommand{\prob}{\Pr}
\spnewtheorem{myclaim}{Claim}{\bfseries}{\itshape}
\newextmathcommand{\NewPsi}{\overline\MyPsi}
\newextmathcommand{\HashedPsi}{\MyPsi_h}
\newextmathcommand{\HashedNewPsi}{\overline\MyPsi_h}
\newextmathcommand{\lastm}{m^*}
\newextmathcommand{\Expect}{\mathsf E}

\journalname{Acta Informatica}

\begin{document}

\title{Approximate Counting in SMT and\\Value Estimation for Probabilistic Programs}
\titlerunning{Approximate Counting in SMT}

\author{
    Dmitry Chistikov
        \and
    Rayna Dimitrova
        \and
    Rupak Majumdar
}

\institute{%
    Max Planck Institute for Software Systems (MPI-SWS), Germany\\
    \email{\{dch,rayna,rupak\}@mpi-sws.org}
}

\date{Received: date / Accepted: date}

\maketitle

\begin{abstract}
\SharpSMT, or model counting for logical theories,
is a well-known hard problem that generalizes such tasks as
counting the number of satisfying assignments to a Boolean formula
and computing the volume of a polytope.
In the realm of satisfiability modulo theories (SMT)
there is a growing need for model counting solvers,
coming from several application domains
(quantitative information flow, static analysis of probabilistic programs).
In this paper, we show a reduction from an approximate
version of \SharpSMT to SMT.

We focus on the theories of integer arithmetic and linear real arithmetic.
We propose model counting algorithms that provide approximate solutions
with formal bounds on the approximation error. They run in polynomial time
and make a polynomial number of queries to the SMT solver for the underlying
theory, exploiting ``for free''
the sophisticated heuristics implemented within modern SMT solvers.
We have implemented the algorithms and used them to solve the value
problem for a model of loop-free probabilistic programs with nondeterminism.
\keywords{
    \SharpSMT \and
    model counting \and
    satisfiability modulo theory \and
    \SharpSAT \and
    volume computation \and
    approximation algorithms \and
    probabilistic programming
}
\end{abstract}


\section{Introduction}

Satisfiability modulo theories (SMT) is a foundational problem in formal methods,
and the research landscape is not only enjoying the success of existing SMT solvers,
but also generating demand for new features. 
In particular, there is a growing 
need for \emph{model counting} solvers; for example,
questions in quantitative information flow and in static analysis of
probabilistic programs are naturally cast as instances of model counting problems
for appropriate logical theories \cite{JhaLICS,SaxenaPLDI14,SCG13}.

We define the \SharpSMT problem that generalizes several model counting
questions relative to logical theories,
such as computing the number of satisfying assignments to a Boolean formula (\SharpSAT)
and computing the volume of a bounded polyhedron in a finite-dimensional real vector space.
Specifically, to define model counting modulo a \df{measured theory},
first suppose every variable in a logical formula comes with a domain which is
also a measure space. Assume that, for every logical formula $\phi$ in the theory,
the set of its models $\allmod\phi$ is measurable with respect to the product
measure; the \df{model counting (or \SharpSMT)} problem then asks, given $\phi$,
to compute the measure of $\allmod\phi$, called the \df{model count} of $\phi$.

In our work we focus on the model counting problems
for the theories of bounded integer arithmetic and linear real arithmetic.
These problems are complete for the complexity class \SharpP, so fast exact
algorithms are unlikely to exist.

We extend to the realm of SMT the well-known hashing approach from the world of \SharpSAT,
which reduces \df{approximate} versions of counting to decision problems.
From a theoretical perspective, we solve a model counting problem with
a resource-bounded algorithm that has access to an oracle
for the decision problem.
From a practical perspective, we show how to use unmodified existing SMT solvers
to obtain approximate solutions to model-counting problems.
This reduces an approximate version of \SharpSMT to SMT.

Specifically, for integer arithmetic (not necessarily linear), we give a randomized algorithm that approximates
the model count of a given formula $\varphi$ to within a multiplicative factor $(1 + \eps)$ for any given $\eps > 0$.
The algorithm makes $O(\frac{1}{\eps} \, | \varphi |)$ SMT queries of size at most
$O(\frac{1}{\eps^2} \, | \varphi |^2)$ where $| \varphi |$ is the size of $\varphi$.

For linear real arithmetic, we give a randomized algorithm that approximates
the model count with an additive error $\gamma N$, where $N$ is the volume
of a box containing all models of the formula, and the coefficient $\gamma$
is part of the input. The number of steps
of the algorithm and the number of SMT queries (modulo the combined theory of
integer and linear real arithmetic) are again polynomial.

As an application, we show how to solve the value
problem (cf.~\cite{SCG13})
for a model of loop-free probabilistic programs with nondeterminism.

\subsubsection*{Techniques}

Approximation of $\SharpP$ functions by randomized algorithms
has a rich history in complexity theory \cite{Stockmeyer1985,VV86,JVV1986,jerrum1996markov}.
Jerrum, Valiant, and Vazirani~\cite{JVV1986} described a hashing-based $\BPP^\NP$\! procedure
to approximately compute any $\SharpP$ function, and noted that this procedure already appeared implicitly in
previous papers by Sipser \cite{Sipser83} and Stockmeyer \cite{Stockmeyer1985}.
The procedure works with encoded computations of a Turing machine
and is thus unlikely to perform well in practice.
Instead, we show a direct reduction from approximate model counting to SMT solving,
which allows us to retain the structure of the original formula.
An alternate approach could eagerly encode \SharpSMT problems into \SharpSAT, 
but experience with SMT solvers suggests that a ``lazy'' approach may be preferable for some problems. 

For the theory of linear real arithmetic, we also need an ingredient
to handle continuous domains. 
Dyer and Frieze~\cite{DyerFrieze1988} suggested
a discretization that introduces bounded additive error; this placed
approximate volume computation for polytopes---or, in logical terms,
approximate model counting for quantifier-free
linear real arithmetic---in \SharpP.
Motivated by the application in the analysis of probabilistic programs,
we extend this technique to 
handle formulas with existentially quantified variables,
while Dyer and Frieze only work with quantifier-free formulas.
To this end, we prove a geometric result that bounds the effect of projections:
this gives us an approximate model counting procedure for existentially quantified
linear arithmetic formulas.
Note that applying quantifier elimination as a preprocessing step can
make the resulting formula exponentially big;
instead, our approach works directly on the original formula that 
contains existentially quantified variables. 

We have implemented our algorithm on top of the Z3 SMT solver~\cite{Z3} and applied it to
formulas that encode the value problem
for probabilistic programs.
Our initial experience suggests that simple randomized algorithms using off-the-shelf SMT solvers can be
reasonably effective.

\subsubsection*{Counting in SMT}

\SharpSMT is a well-known hard problem whose instances have been studied before, e.\,g., in volume computation~\cite{DyerFrieze1988},
in enumeration of lattice points in integer polyhedra~\cite{Barvinok},
and as \SharpSAT~\cite{GomesSS09}.
Indeed, very simple sub-problems, such as counting the number of
satisfying assignments of a Boolean formula or computing the volume of
a union of axis-parallel rectangles in $\R^n$
(called Klee's measure problem~\cite{Klee}) 
are already $\SharpP$-hard
(see Section~\ref{s:theory} below).

Existing techniques for \SharpSMT either
incorporate model counting primitives
into propositional reasoning~\cite{mlz09cade,ZhouHXHCG14,BellePB15}
or are based on enumerative combinatorics~\cite{latte,SaxenaPLDI14,JhaLICS}.
Typically, exact algorithms~\cite{latte,mlz09cade,JhaLICS}
are exponential in the worst case,
whereas approximate algorithms~\cite{SaxenaPLDI14,ZhouHXHCG14}
lack provable performance guarantees.
In contrast to exact counting techniques,
our procedure is easily implementable
and uses ``for free'' the sophisticated heuristics built in off-the-shelf SMT solvers.
Although the solutions it produces are not exact,
they provably meet user-provided requirements on approximation quality.
This is achieved by extending the hashing approach from
SAT~\cite{GomesHSS07,GomesSS09,ChakrabortyMV13cp,ErmonGSS13}
to the SMT context.

A famous result of Dyer, Frieze, and Kannan~\cite{DyerFriezeKannan1991}
states that the volume of a convex polyhedron can be approximated with a multiplicative error
in probabilistic polynomial time (without the need for an SMT solver).
In our application, analysis of probabilistic programs,
we wish to compute the volume of a projection of
a Boolean combination of polyhedra; in general, it is, of course, non-convex.
Thus, we cannot apply the volume estimation algorithm of \cite{DyerFriezeKannan1991},
so we turn to the ``generic'' approximation of $\SharpP$ using an $\NP$ oracle instead.
Our \SharpSMT procedure for linear real arithmetic allows an additive error in the approximation;
it is known that the volume of a polytope does not always have
a small exact representation as a rational number~\cite{Lawrence90}.

An alternative approach to approximate \SharpSMT
is to apply Monte Carlo methods for volume estimation.
They can easily handle complicated measures
for which there is limited symbolic reasoning available.
Like the hashing technique, this approach is also exponential
in the worst case~\cite{jerrum1996markov}:
suppose the volume in question, $p$, is very small
and the required precision is a constant multiple of~$p$. In this case,
Chernoff bound arguments would suggest the need for $\MyOmega(\frac{1}{p})$ samples;
the hashing approach, in contrast, will perform well.
So, while in ``regular'' settings (when $p$ is non-vanishing) the Monte Carlo approach
performs better, ``singular'' settings (when $p$ is close to zero) are better handled by
the hashing approach.
The two techniques, therefore, are complementary to each other
(see the remark at the end of subsection~\ref{s:prog:formulas}).

\subsubsection*{Related work}

Probably closest to our work is a series of papers
by Chakraborty, Meel, Vardi et~al.~\cite{ChakrabortyMV13,ChakrabortyMV13cp,ChakrabortyFMSV14},
who apply the hashing technique to uniformly sample satisfying assignments
of SAT formulas~\cite{ChakrabortyMV13}. They use CryptoMiniSat~\cite{CryptoMiniSat} as a practical implementation
of an \NP (\SAT) oracle, as it has built-in support for XOR (addition modulo~2) constraints
that are used for hashing. Their recent work~\cite{ChakrabortyFMSV14}
supports weighted sampling and weighted model counting,
where different satisfying assignments are associated with possibly different probabilities
(this can be expressed as a discrete case of \SharpSMT).
Concurrently, Ermon et~al.~\cite{ErmonGSS13} apply the hashing technique in the context of
counting problems, relying on CryptoMiniSat as well.
Ermon et~al.\ also consider a weighted setting
where the weights of satisfying assignment are given in a factorized form;
for this setting, as a basic building block,
they invoke an optimization solver ToulBar2~\cite{toulbar2}
to answer MAP (maximum a posteriori assignment) queries.
More recently and concurrently with (the conference version of) our work,
Belle, Van den Broeck, and Passerini~\cite{BelleUAI15}
apply the techniques of Chakraborty et al.\ in the context
of so-called weighted model integration. This is an instance
of \SharpSMT, where the weights of the satisfying assignments (models)
are computed in a more complicated fashion.
Belle et~al.\ adapt the procedure
of Chakraborty et~al., also using CryptoMiniSat, but additionally rely on the Z3
SMT solver to check candidate models against the theory constraints (real arithmetic
in this case) encoded by the propositional variables,
and use the LattE tool~\cite{latte} for computing the volume of polyhedra.

We briefly review the problem settings of Ermon et~al.~\cite{ErmonGSS13}
and Belle et~al.~\cite{BelleUAI15,BellePB15} in Section~\ref{s:theory}.
In our work, the problem setting is more reminiscent of those
in Chakraborty et~al.\ and Ermon et~al. Our implementation currently
does not rely on CryptoMiniSat and uses an unmodified theory solver, Z3,
instead (see subsection~\ref{s:prog:eval}).

\subsubsection*{Contributions}

We extend, from SAT to SMT, the hashing approach to approximate model counting:
\begin{compactenumerate}

\item
We formulate the notion of a measured theory
(Section~\ref{s:theory}) that gives a unified
framework for model-counting problems.

\item
For the theory of bounded integer arithmetic, we provide a direct
reduction (Theorem~\ref{th:mc-integer} in Section~\ref{s:theory}) from
approximate counting to SMT.

\item
For the theory of bounded linear real arithmetic, we give a 
technical construction (Lemma~\ref{l:df-exists} in subsection~\ref{s:approx:cont})
that lets us extend the results of Dyer and Frieze to the case where the polytope
is given as a projection of a Boolean combination of polytopes;
this leads to an approximate model counting procedure for
this theory (Theorem~\ref{th:mc-real} in Section~\ref{s:theory}).

\item As an application, we solve the value problem
for small loop-free probabilistic programs with nondeterminism (Section~\ref{s:prog}).

\end{compactenumerate}

\noindent
The conference version of this paper appeared as~\cite{ChistikovDM15}.

\section{The \SharpSMT Problem}
\label{s:theory}

We present a framework for a uniform treatment of model counting
both in discrete theories like SAT (where it is literally counting models) and in linear
real arithmetic (where it is really volume computation for polyhedra).
We then introduce the notion of approximation and give an algorithm for approximate model
counting by reduction to SMT.

\subsubsection*{Preliminaries: Counting Problems and \SharpP}
%
A relation $R \subseteq \MySigma^* \times \MySigma^*$ is a \df{p-relation}
if (1) there exists a polynomial $p(n)$ such that if $(x,y) \in R$ then $|y| = p(|x|)$
and (2) the predicate $(x,y) \in R$ can be checked in deterministic polynomial time in
the size of $x$.
Intuitively, a p-relation relates inputs $x$ to solutions $y$.
It is easy to see that a decision problem $L$ belongs to $\NP$ if there is a p-relation
$R$ such that $L = \set{x \mid \exists y. R(x,y)}$.

A {\em counting problem} is a function that maps $\MySigma^*$ to $\N$.
A counting problem $f\colon \MySigma^* \rightarrow \N$ belongs to the class
$\SharpP$ if there exists a p-relation $R$ such that 
$
f(x) = | \set{ y \mid R(x,y) } |
$,
i.\,e., the class $\SharpP$ consists of functions that count the number of solutions to a p-relation \cite{Valiant1979}.
\df{Completeness} in \SharpP is with respect to Turing reductions;
the same term is also (ab)used to encompass problems that reduce to a fixed number of queries
to a \SharpP function (see, e.\,g.,~\cite{DyerFrieze1988}).

\SharpSAT is an example of a \SharpP-complete problem: it asks for the number
of satisfying assignments to a Boolean formula in conjunctive normal form (CNF)~\cite{Valiant1979}.
Remarkably, \SharpP characterizes the computational complexity
not only of ``discrete'' problems, but also of problems involving
real-valued variables:
approximate volume computation (with additive error) for bounded rational polyhedra 
in $\R^k$ is \SharpP-complete~\cite{DyerFrieze1988}.

\subsubsection*{Measured Theories
and \SharpSMT}
%
We will now define the notion of model counting
that generalizes \SharpSAT and volume computation for polyhedra.
Suppose \Theory is a logical theory.
Let $\phi(x)$ be a formula in this theory
with free first-order variables $x = (x_1, \ldots, x_k)$.
Assume that \Theory comes with a fixed interpretation
which specifies domains of the variables,
denoted $D_1, \ldots, D_k$, and assigns a meaning to predicates and
function symbols in the signature of \Theory.
Then a tuple $a = (a_1, \ldots, a_k) \in D_1 \times \ldots \times D_k$
is called a \df{model} of $\phi$ if the sentence $\phi(a_1, \ldots, a_k)$
holds, i.\,e., if $a \satisfies_\Theory \phi(x)$. We denote the set of
all models of a formula $\phi(x)$ by $\allmod \phi$;
the \df{satisfiability problem} for \Theory asks, for a formula \phi
given as input, whether $\allmod\phi \ne \emptyset$.

Consider the special cases of \SharpSAT and volume computation for polyhedra;
the corresponding satisfiability problems are \SAT and linear programming.
For \SharpSAT, atomic predicates are of the form $x_i = b$, for $b \in \Bin$,
the domain $D_i$ of each $x_i$ is \Bin,
and formulas are propositional formulas in conjunctive normal form.
For volume computation, atomic predicates are of the form
$c_1 x_{1} + \ldots + c_k x_{k} \le d$, for $c_1, \ldots, c_k, d \in \R$,
the domain $D_i$ of each $x_i$ is $\R$,
and formulas are conjunctions of atomic predicates.
Sets $\allmod\phi$ in these cases are the set of satisfying assignments
and the polyhedron itself, respectively.

Suppose the domains $D_1, \ldots, D_k$ given by the fixed interpretation
are measure spaces: each $D_i$ is associated with a $\sigma$-algebra
$\F_i \sset 2^{D_i}$ and a measure $\mu_i \colon \F_i \to \R$.
This means, by definition, that $\F_i$ and $\mu_i$ satisfy the following properties:
$\F_i$ contains $\emptyset$ and is closed under complement and countable unions,
and $\mu_i$ is non-negative, assigns $0$ to $\emptyset$, and is $\sigma$-additive.%
\footnote{The reader is referred to standard textbooks
on probability and/or measure theory
for further background; see, e.g.,~\cite[Chapter~1]{textbook-probability}.}

In our special cases, these spaces are as follows.
For \SharpSAT, each $\F_i$ is the set of all subsets of $D_i = \Bin$,
and $\mu_i(A)$ is simply the number of elements in $A$.
For volume computation, each $\F_i$ is the set of all Borel subsets of $D_i = \R$,
and $\mu_i$ is the Lebesgue measure.

Assume that each measure $\mu_i$ is $\sigma$-finite, that is,
the domain $D_i$ is a countable union
of measurable sets (i.\,e., of elements of $\F_i$, and so with finite measure associated with them).
This condition, which holds for both special cases, implies that
the Cartesian product $D_1 \times \ldots \times D_k$ is measurable
with respect to a unique \df{product measure} $\mu$, defined as follows.
A set $A \sset D_1 \times \ldots \times D_k$ is \df{measurable} (that is, $\mu$ assigns
a value to $A$) if and only if $A$ is an element of the smallest $\sigma$-algebra
that contains all sets of the form
$A_1 \times \ldots \times A_k$, with $A_i \in \F_i$ for all $i$.
For all such sets,
it holds that $\mu(A_1 \times \ldots \times A_k) = \mu_1(A_1) \ldots \mu_k(A_k)$.

In our special cases, the product measure $\mu(A)$ of a set $A$ is
the number of elements in $A \sset \Bin^k$ and the volume of $A \sset \R^k$,
respectively.

We say that the theory \Theory is \df{measured} if for every formula
$\phi(x)$ in \Theory with free (first-order) variables $x = (x_1, \ldots, x_k)$
the set $\allmod \phi$ is measurable. We define the \df{model count} of
a formula \phi as $\mc\phi = \mu(\allmod \phi)$.
Naturally, if the measures in a measured theory can assume non-integer values,
the model count of a formula is not necessarily an integer.
With every measured theory we associate a \df{model counting problem},
denoted $\SharpSMT[\Theory]$: the input is a logical
formula $\phi(x)$ in \Theory, and the goal is to compute the value $\mc\phi$.

The \SharpSAT and volume computation problems are just special cases
as intended, since $\mc\phi$ is equal to
the number of satisfying assignments of a Boolean formula
and to the volume of a polyhedron, respectively.

Note that one can alternatively restrict the theory to a fixed
number of variables $k$, i.e., to $x = (x_1, \ldots, x_k)$,
where $x \in D_1 \times \ldots \times D_k$, and introduce
a measure $\mu$ directly on $D_1 \times \ldots \times D_k$;
that is, $\mu$ will not be a product measure.
Such measures arise, for instance, when $\mu$ comes in a factorized
form where factors span non-singleton subsets of $\{x_1, \ldots, x_k\}$.
A toy example, with $k = 3$, might have $\mu$ induced by the probability
density function $Z \cdot f_1(x_1, x_2) \cdot f_2(x_2, x_3)$,
where $f_1$ and~$f_2$ are non-negative and absolutely continuous,
and the normalization constant $Z$ (sometimes called \emph{the partition function})
is chosen in such a way that $\mu(D_1 \times D_2 \times D_3) = 1$.
Note that computing~$Z$, given $f_1$ and $f_2$, is itself a \SharpSMT- (i.e., model counting)
question:
the associated theory has measure $\bar \mu$ induced by $f_1 \cdot f_2$,
and the goal is to compute $\mc\True$, where we assume that \True is a formula
in the theory with $\allmod\True = D_1 \times D_2 \times D_3$.
(Much more sophisticated) problems of this form arise in machine learning
and have been studied, e.g., by Ermon et~al.~\cite{ErmonGSS13}.


\begin{remark*}
A different stance on model counting questions,
under the name of weighted model integration (for real arithmetic),
was recently suggested by Belle, Passerini, and Van den Broeck~\cite{BellePB15}.
Their problem setting starts with a tuple
of real-valued (theory) variables~$x = (x_1, \ldots, x_k)$
and a logical formula $\phi$ over~$x$ and over standalone propositional variables, $p = (p_1, \ldots, p_s)$.
All theory atoms in the formula are also abstracted as
(different) propositional variables, $q = (q_1, \ldots, q_t)$.
All literals $l$ of propositional variables $p, q$ are annotated with
weight functions $f_l(x)$, which (can) depend on~$x$.
Take any total assignment to $p, q$ that satisfies the propositional
abstraction of~$\phi$ and let $L$ be the set of all satisfied literals.
The weight of this assignment to $p, q$ is
the integral $\int \prod_{l \in L} f_l(x) \, dx$
taken over the area restricted in $\R^k$ by the conjunction of atoms
that are associated with literals $l \in L$.
The weighted model integral of $\phi$ is then the sum of weights
of all assignments (to $p, q$) that satisfy the propositional abstraction of $\phi$.
\end{remark*}

\noindent
We discuss several other model counting problems
in the following subsection.

\subsubsection*{Approximate Model Counting}
We now introduce {\em approximate} \SharpSMT and show how 
approximate \SharpSMT reduces to SMT.
We need some standard definitions.
For our purposes, a \df{randomized algorithm} is an algorithm that
uses internal coin-tossing. 
We always assume, whenever we use the term, that, for each possible input $x$ to \A,
the overall probability, over the internal coin tosses~$r$,
that \A outputs a wrong answer is at most $1/4$.
(This error probability $1/4$ can be reduced to any smaller $\alpha > 0$,
by taking the median across $O(\log \alpha^{-1})$ independent runs of \A.)

We say that a randomized algorithm \A
\df{approximates} a real-valued functional problem $\C \colon \MySigma^* \to \R$
\df{with an additive error} if \A takes as input
an $x \in \MySigma^*$ and a rational number $\gamma > 0$
and produces
an output $\A(x, \gamma)$ such that
\begin{equation*}
\prb\bigl[|\A(x, \gamma) - \C(x)| \le \gamma\,\UB(x)\bigr] \ge 3/4,
\end{equation*}
where $\UB \colon \MySigma^* \to \R$ is some specific and efficiently computable upper bound
on the absolute value of $\C(x)$, i.\,e., $|\C(x)| \le \UB(x)$, that comes with the problem $\C$.
Similarly, \A
\df{approximates} a (possibly real-valued) functional problem $\C \colon \MySigma^* \to \R$
\df{with a multiplicative error} if \A takes as input
an $x \in \MySigma^*$ and a rational number $\eps > 0$
and produces
an output $\A(x, \eps)$ such that
\begin{equation*}
\prb\bigl[(1+\eps)^{-1} \C(x) \le \A(x, \eps) \le (1+\eps) \,\C(x)\bigr] \ge 3/4.
\end{equation*}
The computation time is usually considered relative to
$|x| + \gamma^{-1}$ or $|x| + \eps^{-1}$, respectively
(note the inverse of the admissible error).
Polynomial-time algorithms that achieve approximations
with a multiplicative error are also known as
fully polynomial-time randomized approximation schemes (FPRAS)~\cite{JVV1986}.

Algorithms can be equipped with \df{oracles} solving auxiliary problems,
with the intuition that an external solver (say, for \SAT) is invoked.
In theoretical considerations,
the definition of the running time of such an algorithm takes into account
the preparation of \df{queries} to the oracle (just as any other computation),
but not the answer to a query---it is returned within a single time step.
Oracles may be defined as solving some specific problems (say, \SAT) as well as
any problems from a class (say, from \NP).
The following result is well-known.

\begin{prop}[generic approximate counting~\cite{JVV1986,Stockmeyer1985}]
\label{p:apx-count}
Let $\C \colon \MySigma^* \rightarrow \N$ be any member of \SharpP.
There exists a polynomial-time randomized algorithm \A which, using an \NP-oracle,
approximates \C with a multiplicative error.
\end{prop}

\noindent
In the rest of this section, we present our results on
the complexity of model counting problems, $\SharpSMT[\Theory]$, for measured theories.
For these problems, we develop randomized polynomial-time approximation
algorithms equipped with oracles, in the flavour of Proposition~\ref{p:apx-count}.
We describe the proof ideas in Section~\ref{s:proofs}, and details are provided in Appendix.
We formally relate model counting and the value problem for
probabilistic programs in Section~\ref{s:prog};
in the implementation, we substitute an appropriate solver for the theory oracle.
We illustrate our approach on an example in Section~\ref{s:example}.

\paragraph{Integer arithmetic.}
By \IA we denote the \df{bounded} version of integer arithmetic:
each free variable $x_i$ of a formula $\phi(x_1, \ldots, x_k)$
comes with a bounded domain $D_i = [a_i, b_i] \sset \Z$,
where $a_i, b_i \in \Z$.
We use the counting measure $|\cdot|\colon A \sset \Z \mapsto |A|$,
so the model count $\mc\phi$ of a formula \phi is the number
of its models.
In the formulas, we allow existential (but not universal) quantifiers at the top level.
The model counting problem for \IA is \SharpP-complete.

\begin{example}\label{ex:ia-mc}
Consider the formula 
\begin{align*}
\varphi(x) &= \exists y \in [1,10].\  (x \geq 1) \land (x \leq 10) \land (2 x + y \leq 6) \\
           &= \exists y.\ (y \ge 1) \land (y \le 10) \land
                                      (x \geq 1) \land (x \leq 10) \land (2 x + y \leq 6)
\end{align*}
in the measured theory \IA.
This formula has one free variable $x$ and one existentially quantified
variable $y$, let's say both with domain $[0,10]$.
It is easy to see that there exist only two values of $x$, $x \geq 1$, for
which there exists a $y \geq 1$ with $2x + y \leq 6$:
these are the integers~$1$ and~$2$.
Hence,~$\mc\varphi = 2$.
\qed
\end{example}

\begin{theorem}
\label{th:mc-integer}
The model counting problem
for \IA
can be approximated with a multiplicative error
by a polynomial-time randomized algorithm
that has oracle access to satisfiability of formulas in \IA.
\end{theorem}

\paragraph{Linear real arithmetic.}
By \RA we denote the \df{bounded} version of linear real arithmetic,
with possible existential (but not universal) quantifiers at the top level.
Each free variable $x_i$ of a formula $\phi(x_1, \ldots, x_k)$
comes with a bounded domain $D_i = [a_i, b_i] \sset \R$,
where $a_i, b_i \in \R$.
The associated measure is the standard Lebesgue measure, and the model
count $\mc\phi$ of a formula \phi is the volume of its set of models.
(Since we consider linear constraints, any quantifier-free formula defines
a finite union of polytopes. It is an easy geometric fact that
its projection on a set of variables will again be a finite union of bounded polytopes.
Thus, existential quantification involves only finite unions.)

\begin{example}\label{ex:ra-mc}
Consider the same formula 
\begin{align*}
\varphi(x) &= \exists y \in [1,10].\  (x \geq 1) \land (x \leq 10) \land (2 x + y \leq 6) \\
           &= \exists y.\ (y \ge 1) \land (y \le 10) \land
                                      (x \geq 1) \land (x \leq 10) \land (2 x + y \leq 6),
\end{align*}
this time in the measured theory \RA, where $x \in \reals$ and $y \in \reals$.
Note that now $\varphi(x)$ is equivalent to $(x \geq 1) \land (x \leq 2.5)$,
and thus $\mc\varphi = 1.5$:
this is the length of the line segment defined by this constraint.
\qed
\end{example}

We denote the combined theory
of (bounded) integer arithmetic and linear real arithmetic by \IARA.
In the model counting problem for \RA,
the a priori upper bound $\UB$ on the solution
is $\prod_{i = 1}^{k} (b_i - a_i)$;
additive approximation of the problem is \SharpP-complete.

\begin{theorem}
\label{th:mc-real}
The model counting problem
for \RA
can be approximated with an additive error
by a polynomial-time randomized algorithm
that has oracle access to satisfiability of formulas in \IARA.
\end{theorem}

\section{Proof Techniques}
\label{s:proofs}

In this section we explain the techniques behind
Theorems~\ref{th:mc-integer} and~\ref{th:mc-real}.
The detailed analysis can be found in Appendix.

\subsection{Approximate discrete model counting}
\label{s:approx:discr}

We now explain the idea behind Theorem~\ref{th:mc-integer}.
Let $\phi(x)$ be an input formula in \IA and let $x = (x_1, \ldots, x_k)$ be the
free variables of $\phi$.
Suppose $M$ is a big enough integer such that
all models of $\phi$ have components not exceeding $M$, i.\,e.,
$\allmod\phi \sset [0, M]^k$.

Our approach to approximating $\mc\phi = |\allmod\phi|$
follows the construction in Jerrum et al.~\cite{JVV1986}, which
builds upon the following observation.
Suppose our goal is to find a value $v$ such that $v \le \mc\phi \le 2 v$,
and we have an oracle \Estimate, for ``Estimate'', answering questions of the form $\mc\phi \qge N$.
Then it is sufficient to make
such queries to \Estimate for $N = N_m = 2^m$, $m = 0, \ldots, k \log (M + 1)$,
and the overall algorithm design is reduced to implementing such
an oracle efficiently.

It turns out that such an implementation can be done with the help of \emph{hashing}.
Suppose that a hash function $h$, taken at random
from some family \Hash, maps elements of $[0, M]^k$ to
$\Bin^m$. If the family \Hash is chosen appropriately, then
each potential model $w$ is mapped by $h$ to, say, $0^m$
with probability $2^{-m}$; moreover, one should expect that
any set $S \sset [0, M]^k$ of size $d$
has roughly $2^{-m} \cdot d$ elements
in $h^{-1}(0^m) = \{ w \in [0, M]^k \mid h(w) = 0^m \}$.
In other words, if $|S| \ge 2^m$, then $S \cap h^{-1}(0^m)$
is non-empty with high probability,
and if $|S| \ll 2^m$, then $S \cap h^{-1}(0^m)$
is empty with high probability.
Distinguishing between empty and non-empty sets is, in turn,
a satisfiability question and, as such,
can be entrusted to the \IA solver.
As a result, we reduced the approximation of the model count of \phi
to a series of satisfiability questions in \IA.

Our algorithm posts these questions as SMT queries of the form
\begin{equation}
\label{eq:query-to-smt}
\phi(x) \land t(x, x') \land (h'(x') = 0^m),
\end{equation}
where $x$ and $x'$ are tuples of integer variables,
each component of $x'$ is either $0$ or $1$,
the formula $t(x, x')$ says that $x'$ is
binary encoding of $x$, and the \IA formula $h'(x') = 0^m$
encodes the computation of the hash function $h$ on 
input $x$.

\let\oldnl\nl
\newcommand{\nonl}{\renewcommand{\nl}{\let\nl\oldnl}}
\begin{algorithm}[t,nofillcomment]
\caption{Approximate model counting for \IA}
\label{algo:discrete}
\KwIn{formula $\varphi(x)$ in \IA}
\KwOut{value $v \in \mathbb{N}$}
\SetKwInOut{Params}{Parameters}
\Params{
\,$\eps \in (0, 1)$, \tcc{ approximation factor}
\nonl $\phantom{\textbf{Parameters:\;}}$ $\alpha \in (0, 1)$, \tcc{ error probability}
\nonl $\phantom{\textbf{Parameters:\;}}$ $a \in \mathbb{N}$ \tcc{ enumeration limit for SMT solver}
}
\nonl Compute values $\lastm, q, p, r$ based on parameters (see text)\;
\lIf{$(e := \chck(\varphi,p+1)) \leq p$}{\Return{$e$}};\label{alg:ln:exact}\\
$\psi(x,x') = \varphi(x)\wedge t(x,x')$\;
$\psi_q(\mathbf{x},\mathbf{x}') = \psi(x^1,x'^1)\wedge\psi(x^2,x'^2)\wedge\ldots\wedge
\psi(x^q,x'^q)$\;
$k' := \text{number of bits in } \mathbf{x}'$\;
\For{$m=1,\dots,m^*$\label{alg:ln:start}}{
$c := 0$; \tcc{ majority vote counter}
\For{$j=1,\ldots,r$}{
\lIf{$\Estimate(\psi_q,k',m,a)$}{$c := c + 1$}
}
\lIf{$c \leq r/2$}{{\bf break};\label{alg:ln:end}}
}
\Return{$\sqrt[q]{a \cdot 2^{m - 0.5}}$}
\end{algorithm}

\begin{algorithm}[nofillcomment]
\caption{Satisfiability ``oracle'' \Estimate}
\label{algo:estimate}
\KwIn{formula $\psi_q(\mathbf{x},\mathbf{x}')$ in \IA; $k',m,a \in \mathbb{N}$}
\KwOut{$\mathit{true}$ or $\mathit{false}$}
\SetKwInOut{Params}{Parameters}
${h'} := \pickh(k',m)$\;
$\psi_{h'}(\mathbf{x},\mathbf{x}') = \psi_q(\mathbf{x},\mathbf{x}') \wedge ({h'}(\mathbf{x}') = 0^{m})$\;
\Return{$(\chck(\psi_{h'},a) \geq a)$} \tcc{ check if $\psi_{h'}$ has at least $a$ models}
\end{algorithm}

Algorithm~\ref{algo:discrete} is the basis of our implementation.
It returns a value $v$ that satisfies
the inequalities $(1 + \eps)^{-1} \mc\phi \le v \le (1 + \eps)\, \mc\phi$
with probability at least $1 - \alpha$. 
Algorithm~\ref{algo:discrete} uses a set of parameters
to discharge small values by enumeration in the SMT solver
(parameters $a, p$) and to query the solver for larger instances (parameters $\lastm,q,r$).
The procedure \Estimate given as Algorithm~\ref{algo:estimate}
asks the SMT solver for \IA to produce $a$ models (for a positive integer parameter $a$) to 
formulas of the form~\eqref{eq:query-to-smt} by calling the procedure $\chck$.

To achieve the required precision with the desired probability, the 
algorithm constructs a conjunction of $q$ copies of the formula
(over disjoint sets of variables),
where the number of copies $q$ is defined as
\begin{equation*}
q = \Big\lceil\frac{1 + 4 \log(\sqrt{a + 1} + 1) - 2 \log a}{2 \log(1+\eps)}\Big\rceil.
\end{equation*}
This results in a formula with $k' = q k \lceil \log (M + 1) \rceil = O(|\phi| / \eps)$~%
binary variables, where $|\phi|$ denotes the size of the original formula $\phi$.
Then, in lines~\ref{alg:ln:start}--\ref{alg:ln:end},
Algorithm~\ref{algo:discrete} performs for each dimension of the hash function
in the range $\{1,\ldots,\lastm\}$  a majority vote over $r$ calls to the
procedure~\Estimate, where the values of~$\lastm$ and $r$ are computed as
follows:
\begin{align*}
\lastm &= \lfloor k' - 2 \log(\sqrt{a + 1} + 1)\rfloor,
&
r &= \left\lceil
    8 \cdot \ln\left(
            \frac{1}{\alpha} \cdot
            \lfloor k' - 2 \log(\sqrt{a + 1} + 1) \rfloor
    \right)
\right\rceil.
\end{align*}
In a practical implementation, early termination of the majority-vote loop is
possible as soon as the number of positive answers given by \Estimate exceeds $r / 2$.

For formulas $\phi$ with up to $p = \lceil (\sqrt{a+1}-1)^{2 / q} \rceil$ models,
Algorithm~\ref{algo:discrete} returns
the exact model count $\mc\phi$ (line~\ref{alg:ln:exact} in Algorithm~\ref{algo:discrete})
computed by the procedure $\chck$, which
repeatedly calls the solver, counting the number of models up to $p+1$.

The values of $\lastm, q, p$, and $r$ used in Algorithm~\ref{algo:discrete},
as well as the choice of the return value $v = \sqrt[q]{a \cdot 2^{m - 0.5}}$,
guarantee its correctness and are
formally derived in Appendix~\ref{s:app:proofs}.

The family of hash functions \Hash used by $\pickh$
in Algorithm~\ref{algo:estimate} needs to satisfy the condition
of \df{pairwise independence}:
for any two distinct vectors $x_1, x_2 \in [0, M]^{k}$
and any two strings $w_1, w_2 \in \Bin^m$,
the probability that a random function $h \in \Hash$
satisfies $h(x_1) = w_1$ and $h(x_2) = w_2$ is equal to $1 / 2^{2 m}$.
There are several constructions for pairwise independent hash functions;
we employ a commonly used family, that
of random XOR constraints
\cite{VV86,BGP00,GomesSS09,ChakrabortyMV13}.
Given $k'$ and $m$, the family contains (in binary encoding) all
functions $h' = (h'_1, \ldots, h'_m) \colon \Bin^{k'} \to \Bin^m$ with
$
h'_i(x_1\ldots, x_{k'}) = a_{i,0} + \sum_{j=1}^{k'} a_{i,j} x_j
$,
where $a_{i,j}\in\set{0,1}$ for all $i$
and $+$ is the XOR operator (addition in $\mathrm{GF}(2)$).
By randomly choosing the coefficients $a_{i,j}$ we get a random hash function from this family.
The size of each query is thus bounded by $O({k'}^2) = O(\frac{1}{\eps^2} | \varphi |^2)$,
where $| \varphi |$ is again the size of the original formula $\varphi$,
and there will be at most $\lastm + 1 \le k' + O(1) = O(\frac{1}{\eps} | \varphi |)$ queries in total.

Note that the entire argument remains valid even if $\phi$ has existentially
quantified variables: queries~\eqref{eq:query-to-smt} retain them as is.
The prefix of existential quantifiers could simply be dropped from~\eqref{eq:query-to-smt},
as searching for models of quantifier-free formulas already captures
existential quantification.
It is important, though, that the model enumeration done by the procedure~$\chck$
in Algorithms~\ref{algo:discrete} and~\ref{algo:estimate}
only count distinct assignments to the \emph{free} variables of~$\phi$ and~$\psi_{h'}$
respectively.

\subsection{Approximate continuous model counting}
\label{s:approx:cont}

In this subsection we explain the idea behind Theorem~\ref{th:mc-real}.
Let \phi be a formula in \RA;
using appropriate scaling, we can assume without loss of generality that
all its variables share the same domain.
Suppose $\allmod\phi \sset [0, M]^k$ and fix some $\gamma$,
with the prospect of finding a value~$v$ that is at most $\eps = \gamma M^k$ away
from $\mc\phi$ (we take $M^k$ as the value of the upper bound~$\UB$ in the definition
of additive approximation).
We show below how to reduce this task of approximate continuous model counting
to additive approximation of a model counting problem for a formula with a discrete
set of possible models,
which, in turn, will be reduced to that of multiplicative approximation.

We first show how to reduce our continuous problem to a discrete one.
Divide the cube $[0, M]^k$ into $s^k$ small cubes with side $\delta$ each,
$\delta = M / s$. For every $y = (y_1, \ldots, y_k) \in \{0, 1, \ldots, s - 1\}^k$,
set $\psi'(y) = 1$ if at least one point
of the cube $C(y) = \{y_j \delta \le x_j \le (y_j + 1)\, \delta, 1 \le j \le k \}$
satisfies \phi; that is, if $C(y) \cap \allmod\phi \ne \emptyset$.

Imagine that we have a formula $\psi$
such that $\psi(y) = \psi'(y)$ for all $y \in \{0, 1, \ldots, s - 1\}^k$,
and let $\psi$ be written in a theory with a uniform measure that
assigns ``weight'' $M / s$ to each point $y_j \in \{0, 1, \ldots, s - 1\}$;
one can think of these weights as coefficients in numerical integration.
From the technique of Dyer and Frieze~\cite[Theorem~2]{DyerFrieze1988}
it follows that for a quantifier-free $\phi$ and an appropriate value of $s$
the inequality $|\mc\psi - \mc\phi| \le \eps / 2$ holds.

Indeed, Dyer and Frieze prove a statement of this form in the context of
volume computation of a polyhedron, defined by a system of inequalities $A x \le b$.
However, they actually show a stronger statement: given a collection of
$m$ hyperplanes in $\R^k$ and a set $[0, M]^k$, an appropriate setting of $s$ will ensure
that out of $s^k$ cubes with side $\delta = M / s$ only a small number $J$
will be \df{cut}, i.\,e., intersected by some hyperplane. More precisely,
if $s = \left\lceil m k^2 M^k / (\eps / 2) \right\rceil$, then
this number $J$ will satisfy the inequality $\delta^k \cdot J \le \eps / 2$.
Thus, the total volume of cut cubes is at most $\eps / 2$,
and so, in our terms, we have $|\mc\psi - \mc\phi| \le \eps/2$ as desired.

However, in our case the formula \phi need not be quantifier-free and may
contain existential quantifiers at the top level. If $\phi(x) = \exists u . \MyPhi(x, u)$ where
$\MyPhi$ is quantifier-free, then the constraints that can ``cut'' the $x$-cubes
are not necessarily inequalities from $\MyPhi$. These constraints can rather arise from
projections of constraints on variables $x$ and, what makes the problem more
difficult, their combinations. However, we are able to prove the following statement:

\begin{lemma}
\label{l:df-exists}
The number $\bar J$ of points $y \in \{0, 1, \ldots, s - 1\}^k$
for which cubes $C(y)$ are cut
satisfies $\bar\delta^k \cdot \bar J \le \eps/2$
if $\bar \delta = M / \bar s$, where
$\bar s = \left\lceil 2^{\overline m + 2 k} k^2 M^k / (\eps / 2) \right\rceil
        = \left\lceil 2^{\overline m + 2 k} k^2 / (\gamma / 2) \right\rceil$
and $\overline m$ is the number of atomic predicates in $\MyPhi$.
\end{lemma}

\begin{proof}
Observe that a cube $C(y)$ is cut if and only if
it is intersected by a hyperplane defined by some predicate in variables~$x$.
Such a predicate does not necessarily come from the formula~$\MyPhi$ itself,
but can arise when a polytope in variables~$(x, u)$ is projected to the space
associated with variables~$x$.
Put differently, each cut cube $C(y)$
has some $d$-dimensional face with $0 \le d \le k - 1$ that ``cuts'' it;
this face is an intersection of $C(y)$ with some affine subspace~$\pi$ in variables~$x$.

Consider this subspace~$\pi$.
It can be, first, the projection of a hyperplane defined in variables~$(x,u)$ by
an atomic predicate in $\MyPhi$ or, second,
the projection of an intersection of several such hyperplanes.
Now note that each predicate in~$(x, u)$ defines exactly one hyperplane;
an intersection of hyperplanes in $(x, u)$ projects
to some specific affine subspace in variables~$x$.
Therefore, each ``cutting'' affine subspace $\pi$ is associated with
a distinct subset of atomic predicates in $\MyPhi$, where, since the domain is bounded,
we count in constraints $0 \le x_j \le M$ as well. This gives us
at most $2^{\overline m + 2 k}$ cutting subspaces, so it remains to apply the result
of Dyer and Frieze with $m = 2^{\overline m + 2 k}$.
\qed
\end{proof}

\noindent
A consequence of the lemma is that the choice of the number $\bar s$ ensures that
the formula
$
\psi(y) = \exists\,x . (\phi(x) \land x \in C(y))
$
written in the combined theory \IARA satisfies the inequality $|\mc\psi - \mc\phi| \le \eps/2$.
Here we associate the domain of each free variable $y_j \in \{0, 1, \ldots, \bar s - 1\}$
with the uniform measure $\mu_j(v) = M / \bar s$. 
Note that the value of $\bar s$ chosen in Lemma~\ref{l:df-exists}
will still keep the number of steps of our algorithm polynomial in the size of the input,
because the number of bits needed to store the integer index along each axis
is $\lceil\log(\bar s + 1)\rceil$ and not $\bar s$ itself.

As a result, it remains to approximate $\mc\psi$ with additive error
of at most $\eps' = \eps/2 = \gamma M^k \!/ 2$, which can be done
by invoking the procedure from Theorem~\ref{th:mc-integer}
that delivers approximation with multiplicative error
$\beta = {\eps'} / {M^k} = \gamma / 2$.

\section{A Fully Worked-Out Example}
\label{s:example}

We now show how our approach
to \SharpSMT, developed in Sections~\ref{s:theory} and~\ref{s:proofs} above,
works on a specific example, coming from
the value problem for \emph{probabilistic programs}.
Probabilistic programs are a means of describing probability distributions;
the model we use combines probabilistic assignments and nondeterministic choice,
making programs more expressive, but analysis problems more difficult.

For this section we choose a relatively high level of presentation in order to
convey the main ideas in a more understandable way;
a formal treatment
follows in Section~\ref{s:prog},
where we discuss (our model of) probabilistic programs and their analysis in detail.

\subsubsection*{The Monty Hall problem~\textup{\cite{selvin75,montyhall}}}
We describe our approach using as an example the following classic problem from
probability theory.
Imagine a television game show with two characters: the player and the host.
The player is facing three doors, numbered $1$, $2$, and $3$;
behind one of these there is a car, and behind the other two there are goats.
The player initially picks one of the doors, say door $i$, but does not
open it. The host, who knows the position of the car, then opens
another door, say door $j$ with $j \ne i$, and shows a goat behind it.
The player then gets to open one of the remaining doors. There are two
available strategies: \emph{stay} with the original choice, door $i$,
or \emph{switch} to the remaining alternative, door $k \not\in \{i, j\}$.
The Monty Hall problem asks, which strategy is better?
It is widely known
that, in the standard probabilistic setting of the problem,
the switching strategy is the better one: it has payoff $2 / 3$,
i.\,e., it chooses the door with the car with probability $2 / 3$;
the staying strategy has payoff of only $1 / 3$.

\subsubsection*{Modeling with a probabilistic program}
We model the setting of the Monty Hall problem with
the probabilistic program in Procedure~\ref{a-proc:mh},
which implements the ``switch'' strategy.
{\LinesNotNumbered
\begin{procedure}[nofillcomment]
\SetProcNameSty{}
\SetKwSwitch{ChooseNonDet}{NonDetOption}{NonDetOther}{choice:}{}{case:}{other:}{}
\caption{3: ``Switch'' strategy in Monty Hall problem()}
\label{a-proc:mh}
$c \sim \Uniform(\{1, 2, 3\})$ \tcc*[f]{position of the car}\\
$i := 1$ \tcc*[f]{initial choice of the player}\\
\ChooseNonDet{}{
    \lNonDetOption{}{ $ j := 2;\ \ASSUME(j \ne c)$\\}
    \lNonDetOption{}{ $ j := 3;\ \ASSUME(j \ne c)$\\}
}
\tcc{the host opens door $j$ with a goat}
\lIf{$i \neq c$}{\ACCEPT} \lElse{\REJECT} \tcc*[f]{the player switches from door $i$}\\
\end{procedure}
}
In this problem, there are several kinds of uncertainty and choice,
so we briefly explain how they are expressed with the features of our
programming model.

First, there is uncertainty in what door hides the car and what
door the player initially picks. It is standard to model the initial
position of the car, $c$, by a random variable distributed uniformly
on $\{1, 2, 3\}$; we simply follow the information-theoretic guidelines here.
At the same time, due to the symmetry of the setting
we can safely assume that the player always picks door $i = 1$ at first,
so here choice is modeled by a deterministic assignment.

Second, there is uncertainty in what door the host opens. We model
this with nondeterministic choice. Since the host knows that the car
is behind door $c$ and does not open door $c$ accordingly,
we restrict this choice by stipulating that $j \ne c$.
For the semantics of the program, this means that for different
outcomes of the probabilistic assignment $c \sim \Uniform(\{1, 2, 3\})$
different sets of paths through the program are available (some
paths are excluded, because they are incompatible with the results
of observations stipulated by \ASSUME statements).

Note that we don't know the nature of the host's choice in the case
that more than one option is available (when $c = 1$, either element
of $\{2, 3\}$ can be chosen as $j$). In principle, this choice may be
cooperative (the host helps the player to win the car),
adversarial (the host wants to prevent the player from winning),
probabilistic (the host tosses a coin), or any other. 
In our example,
the cooperative and the adversarial behavior of the host are identical,
so our model is compatible with either of them.

Finally, uncertainty in the final choice of the player is modeled
by fixing a specific behaviour of the player and declaring acceptance
if the result is successful.
Our procedure implements the ``switching'' strategy; that is, the player always
switches from door $i$. 
The analysis of the program will show how good the strategy is.

\subsubsection*{Semantics and value of the program}
Informally, consider all possible outcomes of the probabilistic assignments.
Restrict attention to those that may result in the program reaching
(nondeterministically) at least one of \ACCEPT or \REJECT statements---such elementary
outcomes form the set \Term (for ``termination''); only these scenarios are
compatible with the observations.
Similarly, some of these outcomes may result in the program reaching (again,
nondeterministically) an \ACCEPT statement---they form the set \Accept; the
interpretation is that for these scenarios the strategy is successful.

These sets \Term and \Accept are events in a probability space.
The \df{value} of the program (in this case interpreted as the payoff
of the player's strategy) is the probability of acceptance conditioned on
termination:
\begin{equation*}
\val{\mathrm{Switch}} =
\prb\,[\Accept \mid \Term] =
\frac{\prb\,[\Accept]}{\prb\,[\Term]},
\end{equation*}
where, in general, we assume $\prb\,[\Term] > 0$ and the last equality follows because $\Accept\cap \Term = \Accept$.
In general, this semantics corresponds to the cooperative behavior of the host,
but in our case the adversarial behavior would be identical: there is no
value of $c$ such that one nondeterministic choice leads to \ACCEPT
and another leads to \REJECT. (We can also deal with adversarial
nondeterminism, see subsection~\ref{s:prog:nondet}.) 
Indeed, for $c = 1$ there are
two paths to \REJECT, and for each of $c = 2, 3$ there is a single
path to \ACCEPT. As a result, $\val{\mathrm{Switch}} = (2 / 3) / (3 / 3) = 2 / 3$,
as intended.

\subsubsection*{Reduction of value estimation to model counting}
To estimate the value of the program, we first reduce its computation
to a model counting problem (as defined in Section~\ref{s:theory})
for an appropriate logical theory.
We write down the verification condition
$\Compute{\exdet, \exprob}$
that defines a valid computation of the program,
by asserting a relation between (values of)
nondeterministic and probabilistic variables \exdet and \exprob.
Then we construct existential formulas of the form
\begin{align*}
\phiacc(\exprob) &= \exists\,\exdet \ . \ \Compute{\exdet, \exprob} \land \ACCEPT \quad\text{and}\\
\phiterm(\exprob) &= \exists\,\exdet \ . \ \Compute{\exdet, \exprob} \land (\ACCEPT \vee \REJECT),
\end{align*}
which assert that the program terminates with ``accept'' (resp.\ ``accept'' or ``reject''),
and whose sets of models (i.\,e., satisfying assignments) are exactly the sets \Accept and \Term
defined above.
For the Monty Hall program, these formulas $\phiacc(c)$ and $\phiterm(c)$,
with $c \in \{1, 2, 3\}$, will be equivalent to $c \neq 1$ and $\extrue$, respectively.
The value of the program is the ratio $\mc\phiacc / \mc\phiterm$, where $\mc\cdot$
denotes the model count of a formula, as in Section~\ref{s:theory}.
Technically, we can use \IA, the theory of integer arithmetic,
with the domain $\{1, 2, 3\}$ for the free variable~$c$ and
with the counting measure $|\cdot|\colon A \mapsto |A|$,
also following Section~\ref{s:theory}.
\label{label:a-example:middle}
So in our example, $\mc\phiacc = 2$ and $\mc\phiterm = 3$.

\subsubsection*{Computing the value of the program}
We show how our method (see subsection~\ref{s:approx:discr}) estimates $\mc\phiacc$.
We make several copies of the variable $c$,
denoted $c^1, \ldots, c^q$. 
The formula
\begin{equation*}
\phi(\exmultc) = \phiacc(c^1) \land \phiacc(c^2) \land \ldots \land \phiacc(c^q)
\end{equation*}
has $2^q$ models, and 
we can estimate $\mc\phiacc$ by estimating $\mc\phi$ and taking the $q$th root of the estimate.
Enlarging $\phiacc$ to $\phi$ and then taking the $q$th root increases precision: for example, 
if the approximation procedure gives a result up to a factor of $2$, the $q$th root of the estimate
for $\mc\phi$ gives an approximation for $\mc\phiacc$ up to a factor of $2^{1 / q}$.

Now observe that for a hash function $h$ with
values in $\{0, 1\}^m$, taken at random from an appropriate family, the expected model count of the formula
\begin{equation}\label{a-eq:query}
\phi(\exmultc) \land (h(\exmultc) = 0^m)
\end{equation}
is $\mc\phi \cdot 2^{-m}$.
By a Chernoff bound argument, the model count is concentrated around the expectation. 
Our algorithm will, for increasing values of $m$,
sample random hash functions from an appropriate class,
construct the formula~\eqref{a-eq:query}, and give the formula to an SMT solver to check satisfiability.
(Note that such formulas are purely existential---in variables $\exmultc$
 as well as in $q$~copies of \exdet.)
With high probability, the first $m$ for which the sampled formula is unsatisfiable
will give a good enough estimate of $\mc\phi$ and, by the reduction above, of $\mc\phiacc$.

\begin{table}
\centering
\small
\caption{Typical run for the Monty Hall example}
\begin{tabular}{lccc}
$m$ & Satisfiable & Unsatisfiable & Majority vote \\
\hline
0\ldots{}6 & 62 & 0 & Sat \\
7\ldots{}9 & 61 & 1 & Sat \\
10  & 55 & 7 & Sat \\
11  & 50 & 12 & Sat \\
12  & 48 & 14 & Sat \\
13  & 21 & 41 & Unsat 
\end{tabular}
\label{a-t:example}
\end{table}

Let us give some concrete values to support the intuition.
We encode the number $c \in \{1, 2, 3\}$ in binary, as $c \equiv c_0 c_1$.
We make $q = 12$ copies, and this will ensure that
we will obtain the \df{exact} value of $\mc\phiacc$
by taking $q$th root of $\mc\phi$, where $\phi$ is as above
(for exact rather than approximate solution,
 a multiplicative gap of less than $3 / 2$ suffices in our setting).
In reality, $\mc\phiacc = 2$ and so $\mc\phi = 2^{12}$, but
we only know a priori that $\mc\phiacc \in [0, 3]$ and $\mc\phi\le 3^{12}$.
We iterate over the dimension $m$ of the hash function and perform the SMT
query~\eqref{a-eq:query} for each $m$.
Using standard statistical techniques, we can reduce the error
probability $\alpha$ by repeating each random experiment
a sufficiently large number of times, $r$; in our case
$r = 62$ leads to $\alpha = 0.01$.
A typical run of our implementation is demonstrated in Table~\ref{a-t:example};
for each $m$ we show how many of the sampled formulas are satisfiable,
and how many are not.
The ``Majority vote'' column is used by our procedure to decide if the number of
models is more than $2^m$ times a constant factor.
From the table, our procedure will conclude that $\mc\phi$ is between $0.17 \cdot 2^{12}$ and $11.66 \cdot 2^{12}$
with probability at least $0.99$
(see Appendix~\ref{s:app:proofs} for derivation of
 the constants $0.17$ and~$11.66$).
This gives us the interval $[1.73,2.45]$ for $\mc\phiacc$;
since $\mc\phiacc$ is integer, we conclude that
$\mc\phiacc = 2$ with probability at least $0.99$.

As mentioned above, the same technique will deliver us $\mc\phiterm = 3$
and hence, $\val{\mathrm{Switch}} = 2 / 3$.

\section{Value Estimation for Probabilistic Programs}
\label{s:prog}

In this section we show how our approach
to \SharpSMT
applies to the \emph{value problem} for probabilistic programs.

\subsubsection*{What are probabilistic programs?}

Probabilistic models such as Bayesian networks, Markov chains, probabilistic guarded-command languages,
and Markov decision processes
have a rich history and form the modeling basis in many different domains (see, e.g., \cite{FV97,McIverMorgan,Darwiche,KollerFriedman}). 
More recently, there has been a move toward integrating probabilistic modeling with ``usual''
programming languages \cite{GTS94,InferNET}.
Semantics and abstract interpretation for probabilistic programs with angelic and demonic non-determinism
has been studied before \cite{Kozen81,McIverMorgan,Monniaux2005,CousotMonerau12}, 
and we base our semantics on these works.

Probabilistic programming models extend ``usual'' nondeterministic programs 
with the ability to sample values from a distribution and condition
the behavior of the programs based on observations \cite{GHNR14}.
Intuitively, probabilistic programs extend an imperative programming language like C with
two constructs: a nondeterministic assignment to a variable from a range of values,
and a probabilistic assignment that sets a variable to a random value sampled from a distribution.
Designed as a modeling framework,
probabilistic programs are typically treated as descriptions of probability distributions
and not meant to be implemented and executed as usual programs.

\subsubsection*{Section summary}

We consider a core \emph{loop-free} imperative language extended with
\emph{probabilistic statements}, similarly to \cite{SCG13}, and with \emph{nondeterministic choice}.
Under each given assignment to the
probabilistic variables, a program accepts (rejects) if there is an execution path
that is compatible with the observations
and goes from the initial vertex to the accepting (resp., rejecting) vertex of its control flow automaton.
Consider all possible outcomes of the probabilistic assignments in a program \program.
Restrict attention to those that result in \program reaching
(nondeterministically) at least one of the accepting or rejecting vertices---such elementary
outcomes form the set \Term (for ``termination''); only these scenarios are
compatible with the observations.
Similarly, some of these outcomes may result in the program reaching (again,
nondeterministically) the accepting vertex---they form the set \Accept.
Note that the sets \Term and \Accept are events in a probability space;
define $\val\program$, the \emph{value} of $\program$, as the conditional probability
$\prb[\Accept \mid \Term]$, which is equal to the ratio
$\frac{\prb[\Accept]}{\prb[\Term]}$ as $\Accept \subseteq \Term$.
We assume that programs are well-formed in that $\prb\,[\Term]$ is bounded away from $0$.

Now consider a probabilistic program \program \df{over a measured theory \Theory,}
i.\,e., where the expressions and predicates come from \Theory.
Associate a separate variable $r$ with each probabilistic assignment in \program
and denote the corresponding distribution by $\dist r$.
Let $R$ be the set of all such variables $r$.

\begin{prop}
\label{p:sa-to-mc}
There exists a polynomial-time algorithm that, given a program \program over \Theory,
constructs logical formulas $\phiacc(R)$ and $\phiterm(R)$ over \Theory
such that $\Accept = \allmod\phiacc$ and $\Term = \allmod\phiterm$,
where each free variable $r \in R$ is interpreted over its domain with measure $\dist r$.
Thus, $\val\program = \mc{\phiacc} / \mc{\phiterm}$.
\end{prop}

\noindent
Proposition~\ref{p:sa-to-mc} reduces the \df{value problem}---i.\,e.,
the problem of computing $\val{\program}$---to model counting.
This enables us to characterize the complexity of the value problem
and solve this problem approximately
using the hashing approach from Section~\ref{s:proofs}.
These results appear as Theorem~\ref{th:value} in subsection~\ref{s:prog:formulas} below.

In the remainder of this section
we define the syntax (subsection~\ref{s:prog:syntax})
and semantics (subsection~\ref{s:prog:semantics}) of our programs
and the value problem.
By reducing this problem
to \SharpSMT (subsection~\ref{s:prog:formulas})
we show an application of our approach to approximate model counting
(an experimental evaluation is provided in subsection~\ref{s:prog:eval}).
We also
discuss modeling different kinds of nondeterminism: cooperative and adversarial
(subsection~\ref{s:prog:nondet}), and
give an short overview of known probabilistic models
subsumed by ours (subsection~\ref{s:prog:examples}).

\subsection{Syntax}
\label{s:prog:syntax}

A program has a set of variables $\vars$,
partitioned into Boolean, integer, 
and real-valued variables.
We assume expressions are type correct, i.e.,
there are no conversions between variables of different types.
The \intextheading{basic statements} of a program are:
\begin{itemize}
\item $\SKIP$ (do nothing),
\item deterministic assignments $x := e$,
\item probabilistic assignments $x\sim \mathsf{Uniform}(a,b)$,
\item assume statements $\ASSUME(\varphi)$,
\end{itemize}
where $e$ and $\varphi$ come from an (unspecified) language of expressions
and predicates, respectively.

The (deterministic) assignment and assume statements have the usual meaning:
the deterministic assignment $x := e$
sets the value of the variable $x$ to the value of the expression
on the right-hand side, and $\ASSUME(\varphi)$ continues execution
only if the predicate is satisfied in the current state
(i.e., it models observations used to condition a distribution).
The probabilistic assignment operation
$x\sim \mathsf{Uniform}(a,b)$
samples the uniform distribution over the range $[a,b]$ with constant parameters $a,b$
and assigns the resulting value to the variable $x$.
For example, for a real variable $x$,
the statement $x \sim \mathsf{Uniform}(0,1)$ draws a value uniformly at random from the segment $[0,1]$,
and for an integer variable $y$, the statement $y\sim\mathsf{Uniform}(0,1)$ sets $y$ to $0$ or $1$ with equal probability.

The \intextheading{control flow} of a program is represented using directed acyclic graphs,
called control flow automata (CFA), whose nodes represent program
locations and whose edges are labeled with program statements.
Let $\mathcal{S}$ denote the set of basic statements; then
a \emph{control flow automaton} (CFA) $\program = (\vars, V, E, \vinit, \vacc, \vrej)$ consists
of a set of variables $\vars$, a labeled, directed, acyclic graph $(V,E)$, with $E \subseteq V \times \mathcal{S} \times V$,
and three designated vertices \vinit, \vacc, and \vrej\ in $V$ called the \emph{initial}, 
\emph{accepting}, and \emph{rejecting} vertices.

\begin{figure}[h]
\begin{center}
\begin{tikzpicture}



\tikzstyle{state} = [circle,draw=black,fill=black,line width = 1pt, inner sep = .07cm, xshift=1.4cm]

\node[state] (s0) at (0,0) {};
\node[state, right of=s0] (s1) {};
\node[state, right of=s1] (s2) {};
\node[state, above right of=s2] (s3) {};
\node[state, below right of=s2] (s4) {};
\node[state, above right of=s4] (s5) {};
\node[state, above right of=s5] (s6) {};
\node[state, below right of=s5] (s7) {};

\node[below of=s0,yshift=.5cm] {$\vinit$};
\node[above of=s6,yshift=-.5cm] {$\vaccept$};
\node[below of=s7,yshift=.5cm] {$\vreject$};

\path[draw = black,line width=1pt] 
(s0) edge[->] node[above,midway,yshift=.3cm]{\small$c \sim \Uniform(\{1, 2, 3\})$} (s1)
(s1) edge[->] node[above,midway]{\small $i := 1$} (s2)
(s2) edge[->] node[above,midway,xshift=-.1cm]{\small $j := 2$} (s3)
(s2) edge[->] node[below,midway,xshift=-.1cm]{\small $j := 3$} (s4)
(s3) edge[->] node[above,midway,xshift=.2cm,yshift=.3cm]{\small $\ASSUME(j \ne c)$} (s5)
(s4) edge[->] node[below,midway,xshift=.2cm,yshift=-.3cm]{\small $\ASSUME(j \ne c)$} (s5)
(s5) edge[->] node[above,midway]{\small $i \neq c$} (s6)
(s5) edge[->] node[below,midway]{\small $i = c$} (s7)
;
\end{tikzpicture}
\end{center}

\caption{CFA for the probabilistic program given as Procedure~\ref{a-proc:mh}.} 
\label{fig:mh-cfa}
\end{figure}

Figure~\ref{fig:mh-cfa} depicts the CFA for the probabilistic program shown in Procedure~\ref{a-proc:mh}.
The \ACCEPT and \REJECT statements from the procedure correspond to the \vaccept and \vreject vertices of the CFA respectively.

We assume \vinit has no incoming edges and \vacc and \vreject have no outgoing edges.
We write $v \xrightarrow{s} v'$ if $(v,s,v') \in E$.
We also assume programs are in static single assignment (SSA) form, that is, 
each variable is assigned at most once along any execution path.
A program can be converted to SSA form using standard techniques \cite{Muchnick,HNRS14}.

Since control flow automata are acyclic, our programs do not have looping constructs. 
Loops can be accommodated in two different ways:
by assuming that the user provides loop invariants \cite{KatoenMMM10}, or
by assuming an outer (statistical) procedure that selects a finite set of executions that
is sufficient for the analysis up to a given confidence level \cite{SCG13,SPMMGC14}.
In either case, the core analysis problem reduces to analyzing finite-path unwindings of programs with loops,
which is exactly what our model captures.

Although our syntax only allows uniform distributions, we can model
some other distributions.
For example, to simulate a Bernoulli random variable $x$ that takes value $0$ with probability
$p$ and $1$ with probability $1-p$, we write the following code:
\begin{align*}
& X \sim \mathsf{Uniform}(0,1); \\
& \mathrm{if}\ (X \leq p) \{ x := 0; \} \mathrm{ else } \{ x:=1; \}
\end{align*}
We can similarly encode uniform distributions with non-constant boundaries
as well as (approximately encode) normal distributions (using repeated samples from uniform distributions 
and the central limit theorem).

\subsection{Semantics}
\label{s:prog:semantics}
The semantics of a probabilistic program is given as a superposition of nondeterministic programs, following \cite{Kozen81,CousotMonerau12}.
Intuitively, when a probabilistic program runs, an oracle makes all random choices faced by the program along its
execution up front.
With these choices, the program reduces to a usual nondeterministic program.

We first provide some intuition behind our semantics.
Let us partition the variables $\vars$ of a program into random variables $R$ (those assigned in a probabilistic assignment) and
nondeterministic variables $N = \vars \setminus R$ (the rest).
(The partition is possible because programs are in static single assignment form.)
We consider two events.
The (normal) \emph{termination} event (resp.\ the \emph{acceptance} event)
states that under a scenario $\omega$ for the random variables in $R$,
there is an assignment to the variables in $N$ such that the program execution under this choice of values
reaches $\vaccept$ or $\vreject$ (resp.\ reaches $\vaccept$).
The termination is ``normal'' in that all $\ASSUME$s are satisfied.
Our semantics computes the conditional probability, under all scenarios, of the acceptance event given that the
termination event occurred.

We now formalize the semantics. 
A \emph{state} of a program is a pair $(v, \bfx)$ of a control node $v\in V$ and a type-preserving assignment 
of values to all program variables in $\vars$.
Let $\MySigma$ denote the set of all states and $\MySigma^*$ the set of finite sequences over $\MySigma$.

Let $(\MyOmega, \F, \prb)$ be the probability space associated with probabilistic assignments in a program \program;
elements of $\MyOmega$ will be called \df{scenarios}.
The probabilistic semantics of $\program$, denoted $\dbrkts{\program}$, is a function from $\MyOmega$ to $2^{\MySigma^*}$,
mapping each scenario $\omega\in\MyOmega$ to a collection of maximal executions of the nondeterministic
program obtained by fixing $\omega$. 
It is defined with the help of an extension of $\dbrkts{\cdot}$ from programs to states,
which, in turn, is defined inductively as follows:
\begin{itemize}
\item
$(\vaccept,\bfx) \in \dbrkts{\vaccept}\omega$ and 
$(\vreject,\bfx) \in \dbrkts{\vreject}\omega$ for all $\bfx$;
\item
$(v,\bfx)(v',\bfx)\sigma \in \dbrkts{v}\omega$ if $v \xrightarrow{\SKIP} v'$
and $(v',\bfx)\sigma \in\dbrkts{v'}\omega$;
\item
$(v,\bfx)(v',\bfx')\sigma \in \dbrkts{v}\omega$
if $v \xrightarrow{x:=e} v'$,
$\bfx' = \bfx[x:= \mathsf{eval}(e)(\bfx,\omega)]$, 
and $(v',\bfx)\sigma \in\dbrkts{v'}\omega$;
similarly, 
if $v \xrightarrow{x\sim \mathsf{Uniform}(a,b)} v'$, we have
$\bfx' = \bfx[x:=c]$ where
$c$ is the value chosen for $x$ in the scenario $\omega$;
\item
$(v,\bfx)(v',\bfx)\sigma \in \dbrkts{v}\omega$
if $v \xrightarrow{\ASSUME(\varphi)} v'$,
$\mathsf{eval}(\varphi)(\bfx, \omega) = \true$, and $(v',\bfx)\sigma \in\dbrkts{v'}\omega$.
\end{itemize}
Finally, define $\dbrkts{\program}\omega = \dbrkts{\vinit}\omega$.
Here $\mathsf{eval}(e)(\bfx, \omega)$ (resp.\ $\mathsf{eval}(\varphi)(\bfx,\omega)$)
denotes the value of the expression $e$ (resp.\ predicate $\varphi$) 
taken in the scenario $\omega$ under the current assignment $\bfx$ of values to program variables,
and $\bfx[x:=c]$ is the assignment that maps variable $x$ to the value $c$ 
and agrees with $\bfx$ on all other variables.

Let $\MyPhi \subseteq \MySigma^*$ be a set of paths of a program \program.
The probability that the run of \program has a property $\MyPhi$ is defined as
\[
\prb\left[\text{run of \program satisfies $\MyPhi$}\right] = \int_{\MyOmega} \1\bigl[\dbrkts{\program} \cap \MyPhi \ne \emptyset\bigr] \, d\prb(\omega)
\]
where $\1\bigl[\dbrkts{\program} \cap \MyPhi \ne \emptyset\bigr]$ denotes the indicator event that
at least one execution path from $\dbrkts\program$ belongs to $\MyPhi$.
Specifically, let $\MyPhi_\acc \subseteq \MySigma^*$ be the set of all sequences that end in a state $(\vaccept,\bfx)$ for some $\bfx$,
and $\MyPhi_\term \subseteq \MySigma^*$ be the set of all sequences that end in either $(\vaccept,\bfx)$ or $(\vreject,\bfx)$.
We define the \emph{termination} and \emph{acceptance} events as 
\begin{align*}
\Term & = \left[\text{run of \program satisfies $\MyPhi_\term$}\right],\\
\Accept & = \left[\text{run of \program satisfies $\MyPhi_\acc$}\right].
\end{align*}
The \emph{value} $\val\program$ of a program $\program$ is defined as the conditional probability
$\prb[\Accept \mid \Term]$, which is equal to the ratio
$\frac{\prb[\Accept]}{\prb[\Term]}$ as $\Accept \subseteq \Term$.
Thus, the value of a program is the conditional probability 
\begin{equation*}
\prb_{\omega} [\,\exists \bfz\,.\,\program(\omega,\bfz) \mbox{ reaches }\vaccept \mid \exists \bfz\,.\,\program(\omega,\bfz) \mbox{ reaches \vaccept or \vreject }\,].
\end{equation*}
For simplicity of exposition, we restrict attention to well-formed programs, 
for which $\prb[\Term]$ is bounded away from $0$.
The \emph{value problem} takes as input a program $\program$ and computes $\val\program$.

Before we show in subsection~\ref{s:prog:formulas} how the value problem
reduces to model counting, we first discuss the features and expressivity of
our model of probabilistic programs. In subsection~\ref{s:prog:nondet} we
discuss the semantics of nondeterminism and in subsection~\ref{s:prog:examples}
we relate our programming model to well-known probabilistic models.

\subsection{Cooperative vs.\ adversarial nondeterminism}
\label{s:prog:nondet}

Our semantics corresponds to a {\em cooperative} understanding of nondeterminism, in the following sense.
For each individual scenario $\omega$, the set $\dbrkts{\program}\omega$ can have one
of the following four forms:
\begin{enumerate}
\renewcommand{\labelenumi}{\theenumi)}
\item\label{a-event:-a-r} there are no paths to \vaccept nor \vreject (for any assignment $\bfz$ for the nondeterministic variables in $N$),
\item\label{a-event:-a+r} there is a path to \vreject, but no paths to \vaccept,
\item\label{a-event:+a-r} there is a path to \vaccept, but no paths to \vreject,
\item\label{a-event:+a+r} there are paths to both \vaccept and \vreject (under different assignments $\bfz,\bfz'$ for the nondeterministic variables).
\end{enumerate}
The conditional probability measure
\begin{equation*}
\prb_\omega [ \ \cdot \mid \Term \,] =
\prb_\omega [ \ \cdot \mid \exists \bfz\,.\,\program(\bfz,\omega) \mbox{ reaches \vaccept or \vreject }\,]
\end{equation*}
restricts the attention to $\omega$ of the forms~\ref{a-event:-a+r}, \ref{a-event:+a-r}, \ref{a-event:+a+r}.
Now our definition of \Accept says that all $\omega$ of the form~\ref{a-event:+a+r} are counted
towards acceptance.
The value of the program is accordingly defined as the (conditional) probability
of options~\ref{a-event:+a-r}, \ref{a-event:+a+r}.

In the Monty Hall problem in Section~\ref{s:example},
this semantics worked as intended only because there are no scenarios $\omega$ of the form~\ref{a-event:+a+r}.
However, a cooperative interpretation may not always be desirable.
Imagine, for instance, that in a game, for some fixed
strategy of the player all scenarios $\omega$ have the form~\ref{a-event:+a+r}, which means
that the outcome of the game depends on the host's choice.
Our semantics evaluates the strategy as perfect, with the value $1$, although
using the strategy may even lead to \emph{losing} with probability $1$ once
nondeterminism is interpreted adversarially.

We can distinguish between semantics with cooperative and adversarial
(also known as angelic and demonic) nondeterminism
by defining the \df{upper} and \df{lower} values of a program by
\begin{align*}
\uval\program & = \prb_{\omega} [\,\exists \bfz\,.\,\program(\bfz,\omega) \mbox{ reaches }\vaccept \mid
          \Term\, ]\quad\text{and}\\
\lval\program & = \prb_{\omega} [\,\nexists \bfz\,.\,\program(\bfz,\omega) \mbox{ reaches }\vreject \mid
          \Term\, ].
\end{align*}
The upper value $\uval\program$ coincides with $\val\program$ as defined in subsection~\ref{s:prog:semantics},
and the lower value $\lval\program$
indeed corresponds to the adversarial interpretation of nondeterministic choice:
only scenarios of the form~\ref{a-event:+a-r} are counted towards acceptance,
and scenarios of the form~\ref{a-event:-a+r} and, most importantly,~\ref{a-event:+a+r}
towards rejection. Obviously, $\lval\program \le \uval\program$, with equality
if and only if the set of scenarios of the form~\ref{a-event:+a+r} has (conditional) measure zero,
as in section~\ref{s:example}.

Observe now that the problem of computing $\lval\program$ reduces to
the problem of computing $\uval\program$:
the reason for that is the equality
\begin{equation*}
\lval\program = 1 - \uval{\dual\program},
\end{equation*}
where for a program $\program = (\vars, V, \allowbreak E, \vinit, \vacc, \vrej)$
we define the corresponding
\df{dual} program $\dual\program = (\vars, V, E, \vinit, \vrej, \vacc)$.
The details are easily checked.

Note that the type of nondeterminism is interpreted at the level of programs and
not on the level of individual statements. Mixing statements with different
type of nondeterminism is equivalent to considering probabilistic programs with
alternation, which raises the complexity of the value problem:
even non-probabilistic loop-free programs with two kinds of nondeterminism on the
per-statement basis are \PSPACE-hard to analyze.

Also note that our semantics resolves the nondeterminism after the probabilistic choice.
This indicates that the nondeterministic choice can ``look in the future.''
For example, consider a program that first chooses a bit $x$ nondeterministically, then
chooses a bit $r$ uniformly at random, and then accepts if $x=r $ and rejects if $x\ne r$.
Under our semantics, the program always accepts: there is a way for the nondeterministic
choice to guess correctly.
This feature of our model can be undesirable in certain cases:
in formal approaches to security, for example, a scheduler
that uses the power to look into the future
when resolving nondeterminism is unrealistic; its existence, however, can
lead to classifying secure protocols as
insecure~\cite{ChatzikokolakisP10}.

We now briefly discuss the {\em synthesis} question in which the nondeterminism
is resolved before the probabilistic choice.
A more general setting, where nondeterministic and probabilistic choice alternate,
is \PSPACE-complete~\cite{Papadimitriou}.

\paragraph{Verification vs.\ synthesis.}
In this paper, we consider the {\em verification} question:
given a probability space over random inputs, the value of the program
is the conditional probability of acceptance, given the program terminates.
As stated above, nondeterminism is resolved
after probabilistic choice.
In decision making under uncertainty, one is also interested in the {\em synthesis}
question: is there a strategy (a way to resolve nondeterministic choices) such
that the resulting probabilistic program achieves a certain value.
That is, the value synthesis problem asks to compute, for a given $p \geq 0$, if 
\[
\exists \bfz \,.\, \val{\program(\cdot, \bfz)} \geq p.
\] 
The complexity of the synthesis problem is, in general, harder than that of the verification
problem. The precise complexity characterization is $\NP^{\SharpP}$, the class of 
problems solvable by a nondeterministic polynomial-time Turing machine with access
to a $\SharpP$ oracle.
Intuitively, the $\NP$-computation guesses the values of variables in $\bfz$, and asks
a $\SharpP$ oracle to resolve the resulting verification problem.
Moreover, the problem is $\NP^{\SharpP}$-hard already for Boolean programs, by using
a reduction from E-MAJSAT, a canonical $\NP^{\SharpP}$-complete problem.

\begin{prop}
Synthesis for probabilistic programs over $\IA$ and $\RA$ is $\NP^{\SharpP}$-complete.
\end{prop}

\noindent
In general, one can study models with arbitrary interleavings of probabilistic and
nondeterministic choice.
For such models, the static analysis problem reduces to \emph{stochastic SMT}, which
is known to be \PSPACE-complete \cite{Papadimitriou}.

We leave the study of ``approximate synthesis'' techniques for the future.

\subsection{Related models}
\label{s:prog:examples}

Our programming model captures (finite-path) behaviors of 
several different probabilistic models that have been
considered before, including the programming models studied
recently \cite{SCG13,HNRS14,SPMMGC14}.
In contrast to models that only capture probabilistic behavior, such as (dynamic) Bayesian networks,
we additionally allow nondeterministic choices.
We show a few additional probabilistic models that can be expressed
as programs.

\paragraph{(Dynamic) Bayesian networks \textup{\cite{Darwiche,KollerFriedman}}.}
A Bayesian network over $V$ is a directed acyclic graph $G = (V,E)$, where each vertex $v\in V$
represents a random variable and each edge $(u,v)\in E$ represents a direct dependence
of the random variable $v$ on the random variable $u$.
Each node $v$ is labeled with a conditional probability distribution:
that of $v$ conditioned on the values of the random variables $\set{u \mid (u,v)\in E}$.
A Bayesian network can be represented as a probabilistic program 
that encodes the conditional probability distribution for each node using a sequence of conditionals
and the Bernoulli distribution.

A temporal graphical model is a probabilistic model for states that evolve over time. 
In such a model, there is a set of random variables $X^{(t)}$ indexed by
a time $t$, and the distribution of a variable $v^{(t+1)} \in X^{(t+1)}$ is given by a conditional
probability distribution over the values of random variables in $X^{(t)}$. 
One example of a temporal model is a dynamic Bayesian network.
A dynamic Bayesian network consists of a pair $\tuple{\mathcal{B}_0,\mathcal{B}_{\rightarrow}}$,
where $\mathcal{B}_0$ is a Bayesian network over $X$ that gives the initial probability distribution
and $\mathcal{B}_{\rightarrow}$ is a Bayesian network over $X\cup X'$, such that only variables 
in $X'$ have incoming edges (or conditional probability distributions associated with them).
Here, $X'$ denotes a fresh copy of variables in $X$.
The network $\mathcal{B}_{\rightarrow}$  defines the distribution of variables in $X'$ 
given values of variables in $X$.
The distribution of $X^{(t+1)}$ is obtained from $X^{(t)}$ according to $\mathcal{B}_{\rightarrow}$.
Given a time horizon $T$, a dynamic Bayesian network is unrolled for $T$ steps in the obvious way:
by first running $\mathcal{B}_0$ and running $T$ copies of $\mathcal{B}_{\rightarrow}$ in sequence. 
Again, for any $T$, such an unrolling can be expressed by a probabilistic program.
Dynamic Bayesian networks subsume several other models, such as hidden Markov models and
linear-Gaussian dynamical systems. 

\paragraph{Influence diagrams \textup{\cite{KollerFriedman}}.} 
Influence diagrams are a common model to study decision making under uncertainty.
They extend Bayesian networks with nondeterministic variables under the control
of an agent.
An influence diagram is a directed acyclic graph $G = (V,E)$, where the nodes are partitioned into
random variables $V_R$, decision variables $V_D$, and utility variables $V_U$. 
Each variable in $V_R \cup V_D$ has a finite domain.
The incoming edges to variables in $V_R$ model direct dependencies as in a Bayesian network,
and the distribution of a random variable is given by a distribution conditioned on the values
of all incoming variables.
Decision variables are chosen by an adversary.
Utility variables have no outgoing edges and model the utility derived by an agent under a given
scenario and choice of decisions.
The value of a utility variable is derived as a deterministic function of values of incoming edges.
For a given scenario of random variables and choice of decision variables, the value of the diagram
is the sum of all utility variables.
By comparing the utility to a constant,
we can reduce computing a bound on the utility to the value problem.
Influence diagrams subsume models such as Markov decision processes with adversarial nondeterminism.%
\footnote{%
	Strictly speaking, MDPs and influence diagrams, where nondeterminism
	is resolved adversarially, are modeled by the \emph{duals} of our programs
	(as defined in subsection~\ref{s:prog:nondet}). 
	Thus, the verification problem asks what is the worst case effect of the environment.
}
The Monty Hall problem in Section~\ref{s:example}
represents an example of an influence diagram.

\paragraph{Probabilistic guarded command languages (pGCL) \textup{\cite{McIverMorgan}}.}
pGCLs extend Dijkstra's guarded command language with a probabilistic choice operation.
They have been used to model communication protocols involving randomization.
Our programs can model bounded unrollings of pGCLs, and the value problem
can be used to check probabilistic assertions of loop-free pGCL code.
This is the core problem in the deductive verification of pGCLs \cite{KatoenMMM10}.

\subsection{From value estimation to model counting}
\label{s:prog:formulas}

We show a reduction from the value problem for a probabilistic program to
a model counting problem.
First, we define a {\em symbolic semantics} of programs.

Let $\program = (\vars, V, E, \vinit, \vaccept, \vreject)$ be a program in
SSA form.
Let $R = \set{ x \in \vars\mid x\sim \mathsf{Uniform}(a,b)\text{ is a statement in }\program}$.
For each variable $r\in R$, we write $\mathsf{dist}(r)$ for the (unique) distribution $\mathsf{Uniform}(a,b)$
such that $r \sim \mathsf{Uniform}(a,b)$ appears in the program.

Let $B_V  = \set{b_v \mid v \in V}$ be a set of fresh Boolean variables.
We associate the following {\em verification condition} $\vc\program$ with the program $\program$:
\begin{align*}
\bigwedge_{v\in V} \left[ b_v \Rightarrow \left( \bigvee_{(v',s,v)\in E} b_{v'} \wedge \MyPsi(s) \right) \right]\wedge b_\vinit
\end{align*}
where $\MyPsi(s)$ is defined as follows:
$\MyPsi(\SKIP)$ is $\true$, $\MyPsi(x:=e)$ is $x = e$, 
$\MyPsi(x\sim\mathsf{Uniform}(a,b))$ is $\true$, and
$\MyPsi(\ASSUME(\varphi))$ is $\varphi$.

Intuitively, the variable $b_v$ encodes ``node $v$ is visited along the current execution.''
The constraints encode that in order for $v$ to be visited, the execution must traverse
an edge $(v', s, v)$ and update the state according to $s$.
The predicate $\MyPsi(s)$ describes the effect of the execution on the state.

The predicates $\MyPsi(s)$ do not add an additional constraint for probabilistic assignments because
we account for such assignments separately as follows.
Define formulas
\begin{align*}
\phiacc  &= \exists B_V \ \exists \vars \setminus R\ .\ \vc\program \wedge b_{\vaccept}, \quad\text{and}\\
\phiterm &= \exists B_V \ \exists \vars \setminus R\ .\ \vc\program \wedge (b_{\vaccept} \vee b_{\vreject}). 
\end{align*}
Note that $\phiacc$ and $\phiterm$ are over the free variables $R$;
if the program \program is \df{over a measured theory \Theory,}
i.\,e., its expressions and predicates come from \Theory,
then \phiacc and \phiterm are formulas in \Theory.

\begin{theorem}[cf. Proposition~\ref{p:sa-to-mc}]
\label{a-th:sa-to-mc}
For a program $\program$, we have $\Accept = \allmod\phiacc$ and $\Term = \allmod\phiterm$,
where each free variable $r \in R$ is interpreted over its domain with measure $\mathsf{dist}(r)$. 
Thus, $\val\program = \mc{\phiacc} / \mc{\phiterm}$.
\end{theorem}

\noindent
Theorem~\ref{a-th:sa-to-mc} reduces the value estimation question to model counting.
Note that our reasoning is program-level as opposed to path-level:
in contrast to other techniques
(see, e.g.,~\cite{SCG13,FilieriPV/13/ReliabilityAnalysis}),
our analysis makes only two \SharpSMT queries and
not one query per path through the program.

For the theories of integer and linear real arithmetic,
Theorem~\ref{a-th:sa-to-mc} gives us a $\SharpP$ upper bound
on the complexity of the value problem.
On the other hand, the value problem is \SharpP-hard, as it easily encodes \SharpSAT.
Indeed, given an instance of $\SharpSAT$ (a Boolean formula in conjunctive normal form), consider 
a program that picks the Boolean variables 
uniformly at random, and accepts iff all the clauses are satisfied.
The number of satisfying assignments to the formula is obtained 
from the probability of reaching the accept vertex. 
Finally,
since the model counting problem can be approximated using a polynomial-time randomized
algorithm with an SMT oracle,
we also get an algorithm for approximate value estimation.

\begin{theorem}[complexity of the value problem]
\label{th:value}
\begin{compactenumerate}
\item 
The value problem for loop-free probabilistic programs (over $\IA$ and $\RA$) is $\SharpP$-complete.
The problem is $\SharpP$-hard even for programs with only Boolean variables.

\item
The value problem for loop-free probabilistic programs over \IA
can be approximated with a multiplicative error
by a polynomial-time randomized algorithm that
has oracle access to satisfiability of formulas in \IA.

\item The value problem for loop-free probabilistic programs over \RA
can be approximated with an additive error
by a polynomial-time randomized algorithm that
has oracle access to satisfiability of formulas in \IARA.

\end{compactenumerate}
\end{theorem}

\paragraph{Remark.}
The core of our value estimation algorithms is a procedure to estimate
the number of models of a formula in a given theory (approximate~\SharpSMT).
An alternative approach to the value problem---and, similarly,
to model counting---would perform Monte Carlo simulation.
It can easily handle complicated probability distributions for which there is limited
symbolic reasoning available.
However, to achieve good performance, Monte Carlo
often depends on heuristics
that sacrifice theoretical guarantees.
In contrast, while using ``for free''
successful heuristics that are already implemented in off-the-shelf SMT solvers to search the state space,
our approach still preserves the theoretical guarantees.

There are simple instances in which Monte Carlo simulation must be run for an exponential
number of steps before providing a non-trivial answer \cite{jerrum1996markov}.
Consider the case when the probability in question, $p$, is very low
and the required precision is a constant multiple of $p$. In such a case,
model counts are small and so there are only a few queries to the SMT solver;
Chernoff bound arguments, in contrast, would suggest running the program $\MyOmega(\frac{1}{p})$ times.

While our SMT-based techniques can also require exponential time within the SMT solver in the worst case,
experience with SMT-based verification of deterministic programs suggests that SMT solvers can be quite
effective in symbolically searching large state spaces in reasonable time.
An illustrative analogy is that the relation between Monte Carlo techniques and SMT-based techniques resembles
that between enumerative techniques
and symbolic techniques in deterministic model checking: while in the worst case, both must enumerate all potential behaviors,
symbolic search often empirically scales to larger state spaces.

In conclusion, Monte Carlo sampling will
easily outperform hashing techniques in a host
of ``regular'' settings, i.\,e., where the probability of termination
is non-vanishing. ``Singular'' settings where this
probability is close to zero will be beyond the reach of Monte Carlo
even for generating a single positive sample (path), let alone
for providing a confidence interval sufficient for multiplicative
approximation of the value of the program. The hashing
approach that we explore deals with such settings easily,
so the two techniques are, in fact, complementary to each other.

\subsection{Evaluation}
\label{s:prog:eval}

We have implemented the algorithm from Subsection~\ref{s:approx:discr} 
in C++ on top of the SMT solver Z3~\cite{Z3}\footnote{More specifically, using version Z3 4.4.0.}. 
The SMT solver is used unmodified, with default settings. 

\subsubsection*{Examples}
\label{s:eval:examples}

We evaluate our techniques on five examples. The first two are probabilistic programs that use nondeterminism.
The remaining examples are Bayesian networks encoded in our language.

\paragraph{The Monty Hall problem\textup{~\cite{selvin75}}}
For the example from Section~\ref{s:example} we compute the probability of
success of the switching strategy.

\paragraph{The three prisoners problem.}
Our second example is a problem that appeared in Martin Gardner's ``Mathematical Games" column in the Scientific American in 1959. 
There, one of three prisoners (1, 2, and 3), who are sentenced to death, is randomly pardoned. 
The guard gives prisoner 1 the following information: If 2 is pardoned, he gives 1 the name of 3. 
If 3 is pardoned, he gives him the name of 2.
If 1 is pardoned, he flips a coin to decide whether to name 2 or 3. 
Provided that the guard tells prisoner 1 that prisoner 2 is to be executed, determine what is prisoner 1's chance to be pardoned?

\paragraph{Pearl's burglar alarm; \;grass model.}
These two examples are classical Bayesian networks from the literature. 
Pearl's burglar alarm example is as given in~\cite[Figure~15]{GHNR14};
the grass model is taken from~\cite[Figure~1]{KiselyovS/09/MonolingualProbabilisticProgramming}.

\paragraph{Kidney disease eGFR sensitivity estimation.}
The last example is a probabilistic model of a medical diagnostics system with noisy inputs. 
We considered the program given in~\cite[Figure~11]{GHNR14} 
using a simplified model of the input distributions. 
In our setting, we approximate the original lognormal distribution 
(the logarithm of the patient's creatinine level) by
drawing its value
uniformly from the set $\{-0.16,-0.09,-0.08,0,0.08,0.09,0.16,0.17\}$, 
regardless of the patient's gender, and we draw the patient's age uniformly from the interval $[30,80]$. 
The patient's gender and ethnicity are distributed in the same way as described in~\cite{SCG13}.

\newcommand{\ccell}[1]{\multicolumn{1}{c|}{\hbox to 2em{\hfil\hss #1 \hss\hfil}}}
\newcommand{\ccelll}[1]{\multicolumn{1}{c||}{\hbox to 2em{\hfil\hss #1 \hss\hfil}}}


\newcommand{\nprobvar}{\textit{free}}
\newcommand{\natom}{\textit{atoms}}

\begin{table} 
\caption{%
    Input and runtime parameters
}
\label{t:eval-in}
\par\medskip
\begin{footnotesize}
\begin{center}
\begin{tabular}{|l||r|r||r|r|r||r|}
\hline
Example &
\ccell{\nprobvar} & 
\ccelll{\natom} & 
\ccell{$\eps$} & 
\ccell{$\alpha$} & 
\ccelll{$a$} & 
\ccell{\rlap{$\,k'$}\phantom{$k$}} \\ \hline
Monty Hall (1)              & $1$ & $5$ & $0.2$      & $0.01$    & $1$  & $24$ \\ \hline
Three prisoners (2)         & $2$ & $6$ & $0.2$      & $0.01$    & $1$  & $36$ \\ \hline
Alarm (3)                   & $4$ & $8$ & $0.5$      & $0.1$     & $20$ & $56$ \\ \hline
Grass model (4)             & $6$ & $8$ & $0.5$      & $0.1$     & $20$ & $48$ \\ \hline
Sensitivity est. (5)        & $8$ & $63$ & $0.5$      & $0.1$     & $20$ & $66$ \\ \hline
\end{tabular}
\par\bigskip
\begin{minipage}{0.75\textwidth}
Legend:
\begin{tabbing}
\quad\=\nprobvar: number of free (probabilistic) variables in the input formula,\\
\>\natom: number of atomic arithmetic predicates in the input formula,\\
\>$\eps$: parameter in the multiplicative approximation factor $(1 + \eps)$,\\
\>$\alpha$: maximum error probability,\\
\>$a$: the SMT enumeration threshold (number of models the SMT solver checks for),\\
\>$k'$: number of binary variables in the formula given to the solver.
\end{tabbing}
\end{minipage}
\end{center}
\end{footnotesize}
\end{table}

\begin{table} 
\caption{%
    Running time of the tool
}
\label{t:eval-out}
\par\medskip
\begin{footnotesize}
\begin{center}
\begin{tabular}{|l||r|r||r|r|}
\hline
Example       & \ccell{$m_\acc$}  & \ccelll{$m_\term$} & time(s) for $\phiacc$  & time(s) for $\phiterm$  \\ \hline
Monty Hall (1)               & $13$  & $20$ & 1.82 & 2.39\\ \hline
Three prisoners (2)          & $0$  & $20$ & 0.01 & 13.54 \\ \hline
Alarm (3)                    & $36$ & $49$ & 130.95 &  80.37\\ \hline
Grass model (4)              & $34$ & $35$ & 52.22 & 53.29 \\ \hline
Sensitivity est. (5)         & $56$ & $57$ & 263.57 & 212.41 \\ \hline
\end{tabular}
\par\bigskip
\begin{minipage}{0.75\textwidth}
Legend:
\begin{tabbing}
$m_\acc$, $m_\term$: maximal hash sizes for $\phiacc$, $\phiterm$, respectively.
\end{tabbing}
\end{minipage}
\end{center}
\end{footnotesize}
\end{table}
\subsubsection*{Results}
\label{s:eval:res}

For each program \program, we used our tool to estimate
the model count of the formulas $\phiacc$ and $\phiterm$;
the value $\val\program$ of the program is approximated
by $v_\acc / v_\term$, where $v_\acc$ and $v_\term$
are the approximate model counts computed by our tool.
Table~\ref{t:eval-in} shows input and runtime parameters for the considered
examples.
The approximation factor $\eps$, the bound $\alpha$ on the error probability, and
the enumeration limit $a$ for the SMT solver are provided by the user. 
For examples (1) and (2), we choose $\eps$ to be $0.2$, while for the remaining examples we take $0.5$.
The chosen value of $\eps$ has impact on the number of copies $q$ of the formula that we construct,
an thus on the number $k'$ of binary variables in the formula given to the solver. 
Furthermore, the more satisfying assignments a formula has, the larger dimension~$m$
of the hash function is reached during the run.
Table~\ref{t:eval-out} shows $m_\acc$ and $m_\term$:
the maximal values of $m$
reached during the runs on $\phiacc$ and
$\phiterm$; it also shows the time (in seconds) our tool
takes to compute $v_\acc$ and $v_\term$.

While our technique can solve these small instances 
in reasonable time, there remains much room for improvement.
Although SAT solvers can scale to large instances, it is well known that even
a small number of XOR constraints 
can quickly exceed the capabilities of state-of-the-art solvers%
~\cite{Urquhart/87/HardExamplesForResolution,SoosNC09,HanJ12}. 
Since for each $m$ we add $m$ parity constraints to the formula, we run into the SAT bottleneck:
computing an approximation of $\mc\phiacc$ for example~(4) with
$\eps = 0.3$ results in running time of several hours.
(At the same time, exact counting by enumerating satisfying assignments
is not a feasible alternative either:
for the formula $\phiacc$ in example~(4), which has more than $400\,000$ of them,
performing this task naively with Z3 also took several hours.)
Our current implementation pre-solves the system of XOR constraints
before passing them to Z3, which somewhat improves the performance;
however, the efficiency of the hashing approach can benefit greatly
from better handling of XOR constraints in the SMT solver.
For example, a SAT solver that deals with XOR constraints efficiently%
---such as Crypto\-Mini\-Sat~\cite{CryptoMiniSat,Soos10}---%
can scale to over a thousand variables~%
\cite{ChakrabortyMV13,ChakrabortyFMSV14,GomesSS09};
incorporating such a SAT solver within Z3 remains a task for the future.
(Needless to say, other families of pairwise independent hash functions can be
 used instead of XOR constraints, but essentially all of them seem to use
 arithmetic modulo~$p$ for $p \ge 2$, which appears hard for theory solvers.)

The scalability needs improvement also in the continuous case,
where our discretization procedure introduces a large number of discrete variables.
For instance, a more realistic model of example (5) would be one in which the logarithm of the creatinine level is 
modeled as a continuous random variable. 
This would result, after discretization, in formulas with hundreds of Boolean variables,
which appears to be beyond the limit of Z3's XOR reasoning.

\section{Concluding Remarks}
\label{s:disc}

Static reasoning questions for probabilistic programs~\cite{GHNR14,SCG13,HNRS14},
as well as quantitative and probabilistic analysis of
software~\cite{BFdPV14,JhaLICS,FilieriPV/13/ReliabilityAnalysis,LuckowPDFV14},
have received a lot of recent attention.
There are two predominant approaches to these questions.
The first one is to perform Monte Carlo sampling
of the program~\cite{SCG13,BFdPV14,LuckowPDFV14,ChagantyNR13,SPMMGC14}.
To improve performance, such methods use sophisticated heuristics
and variance reduction techniques,
such as stratified sampling in~\cite{SCG13,BFdPV14}.
The second approach is based on reduction
to model counting~\cite{FilieriPV/13/ReliabilityAnalysis,JhaLICS,mlz09cade,SaxenaPLDI14},
either using off-the-shelf \SharpSMT solvers
or developing \SharpSMT procedures on top of existing tools.
Another recent approach is based on data flow analysis~\cite{ClaretRNGB13}.
Our work introduces a new dimension of approximation to this area:
we reduce program analysis to \SharpSMT, but
carry out a randomized approximation procedure for the count.
In contrast to previous techniques, our analysis is performed at the program level
and not at the path level: the entire analysis makes only two queries to a \SharpSMT oracle
(not one query per path through the program).

By known connections between counting and uniform generation~\cite{JVV1986,BGP00},
our techniques can be adapted to generate (approximately)
uniform random samples from the set of models of a formula in \IA or \RA.
Uniform generation from Boolean formulas using hashing techniques
was recently implemented and evaluated in the context of constrained random testing
of hardware~\cite{ChakrabortyMV13,ChakrabortyFMSV14}.
We extend this technique to the SMT setting,
which was left as a future direction in \cite{ChakrabortyMV13}
(previously known methods for counting integral points
of polytopes~\cite{Barvinok,JhaLICS} do not generalize to the nonlinear theory \IA).

\subsubsection*{Further directions}

\paragraph{Scalability.}
An extension of the presented techniques may be desirable to cope with
larger instances of \SharpSMT. As argued in subsection~\ref{s:prog:eval},
incorporating XOR-aware reasoning into an SMT solver can be
an important step in this direction.

\paragraph{Theories.}
Similar techniques apply to theories other than \IA and \RA.
For example, our algorithm can be extended to
an appropriate fragment of the combined theory
of string constraints and integer arithmetic.
While SMT solvers can handle this theory (using heuristics), it would be nontrivial
to design a model counting procedure
using the previously known approach based on generating
functions~\cite{SaxenaPLDI14}. 

\paragraph{Distributions.}
Although the syntax of our probabilistic programs
supports only $\Uniform$, it is easy to simulate
other distributions:
Bernoulli,
uniform with non-constant endpoints,
(approximation of) normal.
This, however, will not scale well,
so future work may incorporate
non-uniform distributions as a basic primitive.
(An important special case covers
weighted model counting in \SAT,
for which a novel extension of the hashing approach
was recently proposed~\cite{ChakrabortyFMSV14}
and, by the time the present paper was submitted,
also studied in the context of SMT~\cite{BelleUAI15}.)

\paragraph{Applications.}
A natural application of the uniform generation technique
in the SMT setting would be a procedure that generates
program behaviors uniformly
at random from the space of possible behaviors.
(For the model we studied, program behaviors are trees: the branching comes from
nondeterministic choice, and the random variables are sampled from their respective
distributions.)

\bibliographystyle{spmpsci} 
\bibliography{main}

\newpage
\appendix
\section{Appendix: Technical proofs}
\label{s:app:proofs}

In this section we fill in the details in the proof of Theorem~\ref{th:mc-integer},
continuing subsection~\ref{s:approx:discr} and thus proving correctness
of Algorithm~\ref{algo:discrete};
however, to simplify notation, we write $n$ instead of $k'$ to denote
the total number of Boolean variables.
As the entire analysis is essentially Boolean, we build on
a previous exposition of the topic due to Trevisan~\cite{TrevisanCounting}.
We pay much more attention to the precise choice of parameters, though;
we assume that the SMT enumeration threshold $a \ge 1$,
the approximation parameter $\eps > 0$, and
the upper bound on the probability of bad estimate $\alpha \in (0; 1)$ are given
as input.
In subsection~\ref{s:app:copies} we show how to choose:
\begin{itemize}
\item $q$, the number of copies of the formula,
see equation~\eqref{eq:choice-copies}, and
\item $v$, the output value of the algorithm (during its run),
see equation~\eqref{eq:choice-answer}.
\end{itemize}
In subsection~\ref{s:app:majority} we show
\begin{itemize}
\item $p$, the initial enumeration threshold (number of models that
are sought before the algorithm enters the main loop), see equation~\eqref{eq:choice-initial-threshold},
\item \lastm, the largest possible dimension of the hash,
see equation~\eqref{eq:choice-maxhash}, and
\item $r$, the number of calls to the ``Estimate'' oracle (\Estimate),
see equation~\eqref{eq:choice-majority}.
\end{itemize}
Before this, in subsection~\ref{s:app:estimate}, we show how to choose
internal parameters
(equation~\eqref{eq:param-gap}) so as to
establish key properties of the oracle \Estimate;
the choice of parameters in the following subsections~\ref{s:app:copies}
and~\ref{s:app:majority} relies on these properties.

\subsection{The ``Estimate'' oracle from subsection~\ref{s:approx:discr}}
\label{s:app:estimate}

We use a simple form of the \df{Leftover Hash Lemma}.
This lemma was originally proved by Impagliazzo, Levin, and Luby~\cite{ImpagliazzoLL89};
we use a formulation due to Trevisan~\cite{TrevisanCounting}.
In brief, the lemma establishes the following property:
For any sufficiently large set $S \sset \Bin^n$,
the number of elements of $S$ that are mapped to a particular image, $0^m$,
by a random hash function $h$ from an appropriate family~$\Hash$
does not deviate a lot from its expected value, $|S| / 2^m$.

\begin{lemma}
\label{l:lhl}
Let $\Hash$ be a family of pairwise independent hash functions
$h \colon \Bin^n \to \Bin^m$.
Let $S \sset \Bin^n$ be such that $|S| \ge 4 / \rho^2 \cdot 2^m$.
For $h \in \Hash$, let $\zerothsize$ be the cardinality of
the set $\{ w \in S \colon h(w) = 0^m \}$.
Then
\begin{equation*}
\prob
\left[
\left|\zerothsize - \frac{|S|}{2^m}\right|
\ge
\rho \cdot \frac{|S|}{2^m}
\right]
\le \frac{1}{4}.
\end{equation*}
\end{lemma}

\noindent
We now show how to implement the ``Estimate'' oracle \Estimate.
Recall that its goal, roughly speaking, is to answer questions of the form
\begin{equation}
\label{eq:estimate-question}
\text{Does the formula $\MyPsi$ have at least $N = 2^m$ models?}
\end{equation}
Let $a$ be a positive integer parameter, to be chosen arbitrarily.
Our oracle \Estimate will rely, in turn, on an SMT oracle (solver) for the underlying
theory (\IA) and will post queries of the form
\begin{equation}
\label{eq:smt-enumerate-query}
\text{Does the formula $\HashedPsi := (\MyPsi \land (h = 0))$ have at least $a$~models?}
\end{equation}
where $h$ is a hash function $h \colon \Bin^n \to \Bin^m$
with $m$ chosen in an appropriate way.
Instead of answering questions~\eqref{eq:estimate-question} exactly,
our oracle \Estimate will have a blind spot.
Let $\mc\MyPsi$ be the number of models of $\MyPsi$;
for some parameters $g < G$ and for any sufficiently large $m$,
we ensure that the following properties hold:
(I) if $\mc\MyPsi < g \cdot 2^m$, then \Estimate returns ``no''
with high probability;
(II) if $\mc\MyPsi > G \cdot 2^m$, then \Estimate returns ``yes''
with high probability.
The blind spot is the intermediate case, $g \cdot 2^m \le \mc\MyPsi \le G \cdot 2^m$:
the oracle \Estimate can answer ``yes'' or ``no'' in an arbitrary way.
The entire implementation of the oracle will be very simple:
it will pick $h$ at random from $\Hash$, ask the question~\eqref{eq:smt-enumerate-query}
for the obtained formula $\HashedPsi = (\MyPsi \land (h = 0))$
and repeat the answer---yes or no---of the underlying SMT oracle.

Let us now proceed to proofs of properties~(I) and~(II).

\begin{myclaim}
\label{c:estimate-no}
Let $x > 0$ be a real number such that $g = 4 / x^2$ and $a = (1 + x)\,g$.
Suppose $\mc\MyPsi \le g \cdot 2^m$;
then $\Prob[\text{\Estimate = ``no''}] \ge 3 / 4$.
\end{myclaim}

\begin{proof}
Denote $\theta = \mc\MyPsi / 2^m \le g$
and pick any formula \NewPsi such that, first,
$\allmod\MyPsi \sset \allmod\NewPsi$ and, second,
$\mc\NewPsi = g \cdot 2^m$.
Write $S = \allmod\NewPsi$ and, as above,
let $\zerothsize$ be the cardinality of
the set $\{ w \in S \colon h(w) = 0^m \}$;
note that $\zerothsize = \mc\HashedNewPsi$
where $\HashedNewPsi = \NewPsi \land (h = 0)$.
Observe that
$\Prob[\text{\Estimate = ``yes''}]$
is equal to
\begin{multline*}
\Prob[\mc\HashedPsi - g \ge a - g] \le
\Prob[\mc\HashedNewPsi - g \ge a - g] =
\Prob[\zerothsize - g \ge x \cdot g] \le
\Prob[|\zerothsize - g| \ge x \cdot g] \le 1/4,
\end{multline*}
where the last inequality follows from Lemma~\ref{l:lhl}
with $\rho = x$, since $|S| = \mc\NewPsi = g \cdot 2^m = 4 / x^2 \cdot 2^m$.
\qed
\end{proof}

\begin{myclaim}
\label{c:estimate-yes}
Let $y > 0$ be a real number such that $G = 4 / y^2$ and $a = (1 - y)\,G$.
Suppose $\mc\MyPsi \ge G \cdot 2^m$;
then $\Prob[\text{\Estimate = ``yes''}] \ge 3 / 4$.
\end{myclaim}

\begin{proof}
As in Claim~\ref{c:estimate-no},
denote $\theta = \mc\MyPsi / 2^m \ge G$.
Now pick $S = \allmod\MyPsi$ and
let $\zerothsize$ again be the cardinality of
the set $\{ w \in S \colon h(w) = 0^m \}$;
we have $\zerothsize = \mc\HashedPsi$.
Observe that
$\Prob[\text{\Estimate = ``yes''}]$
is equal to
\begin{multline*}
\Prob[\mc\HashedPsi - \theta \ge a - \theta] \ge
\Prob[\mc\HashedPsi - \theta \ge (1 - y) \cdot G - G] =
\Prob[\zerothsize - \theta \ge - y \cdot G] \ge\\
\Prob[|\zerothsize - \theta| \le y \cdot G] \ge 3/4,
\end{multline*}
where the last inequality again follows from Lemma~\ref{l:lhl},
now with $\rho = y$ and $|S| = \mc\MyPsi = \theta \cdot 2^m \ge G \cdot 2^m = 4 / y^2 \cdot 2^m$.
\qed
\end{proof}

\noindent
Let us match the parameter settings from Claims~\ref{c:estimate-no} and~\ref{c:estimate-yes}.
We have
\begin{align*}
g(x) &= 4 / x^2, & G(y) &= 4 / y^2, \\
a(x) &= (1 + x) \cdot 4 / x^2, &
a(y) &= (1 - y) \cdot 4 / y^2.
\end{align*}
Needless to say, the following equality needs to be satisfied:
\begin{equation}
\label{eq:same-a}
a = a(x) = a(y).
\end{equation}
The multiplicative gap between $G$ and $g$ is
$B = G(y) / g(x) = (x / y)^2$,
i.e., $B = \lambda^2$ for $\lambda = x / y$.
Suppose $\lambda \ge 1$ is fixed; then equation~\eqref{eq:same-a}
gives us
\begin{gather*}
\frac{1 + \lambda y}{\lambda^2 y^2} = \frac{1 - y}{y^2}, \\
1 + \lambda y = \lambda^2 - \lambda^2 y, \\
(\lambda^2 + \lambda) y = (\lambda^2 - 1), \\
y = \frac{\lambda - 1}{\lambda} = 1 - \frac{1}{\lambda} \quad\text{and}\\
a = a(\lambda) = \frac{1}{\lambda} \cdot 4 \cdot \frac{\lambda^2}{(\lambda - 1)^2} =
\frac{4 \lambda}{(\lambda - 1)^2}.
\end{gather*}
Given an integer $a \ge 1$, how big a gap $B(a)$ does it correspond to?
Rewrite the equation $a(\lambda) = a$ as
$(\lambda - 1)^2 \cdot a = 4 \lambda$ and further as
\begin{equation*}
\lambda^2 \cdot a - \lambda \cdot (2 a + 4) + a = 0.
\end{equation*}
Both roots of this quadratic equation are real, but only
the greater one is $\ge 1$; it is given by the formula
\begin{align*}
\lambda(a) &=
\frac{a + 2 + 2 \sqrt{a + 1}}{a} =
\frac{(\sqrt{a + 1} + 1)^2}{a}
\quad\text{and corresponds to}\\
y(a) &= \frac{\lambda(a) - 1}{\lambda(a)} =
\frac{2 + 2 \sqrt{a + 1}}{a + 2 + 2 \sqrt{a + 1}} =
\frac{2}{\sqrt{a + 1} + 1}
\quad\text{and}\\
x(a) &= \lambda(a) \cdot y(a) = \lambda(a) - 1 = \frac{2 \, (\sqrt{a + 1} + 1)}{a}.
\end{align*}
Finally,
\begin{equation}
\label{eq:param-gap}
\begin{aligned}
g &= g(x) = \frac{a^2}{(\sqrt{a + 1} + 1)^2} = (\sqrt{a + 1} - 1)^2, \\
G &= G(y) = (\sqrt{a + 1} + 1)^2,
\quad\text{and}\\
B &= \lambda^2 = \frac{(\sqrt{a + 1} + 1)^4}{a^2} =
\left(\frac{\sqrt{a + 1} + 1}{\sqrt{a + 1} - 1}\right)^2.
\end{aligned}
\end{equation}
To sum up, fixing $a \ge 1$ for SMT queries~\eqref{eq:smt-enumerate-query}
leads to the multiplicative blind spot of ``size'' $B$
and constants~$g$ and~$G$ defined in Equation~\eqref{eq:param-gap};
we will use these parameters in the following subsections.

\subsection{Copies of the formula and return value of the algorithm}
\label{s:app:copies}

Recall that the formula $\MyPsi$ that we use throughout
the algorithm is of the following form:
\begin{equation*}
\MyPsi = \psi_q = \psi^{(1)} \land \psi^{(2)} \land \ldots \land \psi^{(q)}
\end{equation*}
where $q \ge 1$ is a natural parameter and formulas $\psi^{(i)}$, $1 \le i \le q$,
are copies of $\psi$ where all variables are replaced by fresh copies.
In total, $\MyPsi$ has $q$ times as many variables as $\psi$,
and $\mc\MyPsi = \mc\psi^q$.

We now describe how the parameter $q$ is chosen.
Recall that Algorithm~\ref{algo:discrete} calls the ``Estimate'' oracle \Estimate
with $m = 1, 2, \ldots$
(we ignore the majority vote machinery for now; the reader can safely assume $r = 1$
 as the reasoning in the general case is the same).
Suppose first several calls to \Estimate result in ``yes'' answers,
and let $m$ be the first dimension of the hash
that corresponds to a ``no''.
We can now rule out (here and below---with high probability)
the case that $\mc\MyPsi \ge G \cdot 2^m$;
similarly, it is unlikely that $\mc\MyPsi \le g \cdot 2^{m - 1}$.
We should conclude, therefore, that
\begin{equation}
\label{eq:segment-estimate}
g / 2 \cdot 2^m
<
\mc\MyPsi
<
G \cdot 2^m
\end{equation}
with high probability.
Now, the task of the overall algorithm is to return a value $v$ that
lies in the segment $((1 + \eps)^{-1} \cdot \mc\psi; (1 + \eps) \cdot \mc\psi)$;
in other words, $v^q$---which is an estimate of $\mc\MyPsi$---should
satisfy the condition
\begin{equation}
\label{eq:segment-estimate-eps}
(1 + \eps)^{-q} \cdot \mc\MyPsi
\le
v^q
\le
(1 + \eps)^q \cdot \mc\MyPsi.
\end{equation}
We now align the segments defined in equations~\eqref{eq:segment-estimate}
and~\eqref{eq:segment-estimate-eps} above.
First, observe that the ratios of the right and left endpoints for each of these segments
are $2 G / g = 2 B$ and $(1 + \eps)^{2 q}$, respectively;
recall that $B = G / g$ is given by equation~\eqref{eq:param-gap}.
As our goal is thus to ensure that $(1 + \eps)^{2 q} \ge 2 B$,
we choose
\begin{equation}
\label{eq:choice-copies}
q = \left\lceil \frac{1 + \log B}{2 \log (1 + \eps)} \right\rceil.
\end{equation}
Second, our best estimate for $\mc\MyPsi$ is, accordingly, the geometric mean of
the left and right endpoints of the segment in~\eqref{eq:segment-estimate};
in other words, the best estimate for $\log\mc\MyPsi$ is the arithmetic mean
of their logarithms:
\begin{equation*}
\log (v^q) = m + \frac{\log g - 1 + \log G}{2} = m + \frac{\log (g \cdot G) - 1}{2} = \\
m + \frac{\log(a^2) - 1}{2} = m + \log a - \frac{1}{2},
\end{equation*}
and thus the return value of Algorithm~\ref{algo:discrete} is
\begin{equation}
\label{eq:choice-answer}
v = \sqrt[q]{a \cdot 2^{m - 0.5}}.
\end{equation}

\subsection{Majority vote and confidence level}
\label{s:app:majority}

It remains to choose the parameter $r$ that determines
how many times the ``Estimate'' oracle \Estimate is called
for each value of~$m$. The choice of $r$ depends primarily
on $\alpha \in (0; 1)$, a number provided as part of the input:
the probability that the algorithm returns a value~$v$ that
is not within a $(1 + \eps)$-factor of $\mc\psi$ should be
at most~$\alpha$.
For this choice, however, we also take into account the smallest
and largest values of $m$ that can be reached during the run of the algorithm.

We first look into the smallest $m$ on the run.
To make the algorithm simple, we start from $m = 1$;
to achieve the same quality of the produced values as in the previous
subsection, we need to ensure that the maximum possible gap
that arises if \Estimate returns ``no'' for $m = 1$
is (at most) of the same size.
For larger values of~$m$, the algorithm would conclude
that equation~\eqref{eq:segment-estimate} is satisfied.
For $m = 1$, this means that the case $\mc\MyPsi \le g$ should be
ruled out. Therefore, the \emph{initial} enumeration threshold
for $\mc\MyPsi$ should be set to~$g$ (note that, in fact, $g \le a$);
if the enumeration is done on the original formula~$\psi$ instead,
the threshold is chosen as
\begin{equation}
\label{eq:choice-initial-threshold}
p = \lceil g^{1 / q} \rceil = \left\lceil (\sqrt{a + 1} - 1)^{2 / q} \right\rceil
\end{equation}
where $q$ is as in equation~\eqref{eq:choice-copies}.

Let us now look into the largest $m$ on the run.
Here the purpose of the call to \Estimate is essentially to rule
out the case $\mc\MyPsi \ge G \cdot 2^m$; this is needed
unless $G \cdot 2^m > 2^n$ where $n$ is the total number of
Boolean variables. Hence, the last call to \Estimate
should have $m \le \lastm$ where \lastm is defined as
\begin{equation}
\label{eq:choice-maxhash}
\lastm = \lfloor n - \log G \rfloor = \lfloor n - 2 \log(\sqrt{a + 1} + 1)\rfloor.
\end{equation}

Finally, let us proceed to the choice of~$r$. Suppose that $m$
is fixed, and let $X_i$ denote the Bernoulli random variable
that is equal to~$1$ if the $i$th call to \Estimate returns
the \emph{less likely} answer
and to $0$ otherwise
(recall that, by Claims~\ref{c:estimate-no} and~\ref{c:estimate-yes},
 the \emph{less likely} answer is ``no'' if $\mc\MyPsi \ge G \cdot 2^m$
 and ``yes'' if $\mc\MyPsi \le g \cdot 2^m$;
 these answers correspond to \Estimate being ``wrong'').
Let $\Expect$ denote the expectation;
we have $\Expect X_i \le 1/4$ for $1 \le i \le r$ by the choice
of parameters in subsection~\ref{s:app:estimate}.
Denote by $A_m$ the event that, in the presence of $r$ ``voters'',
the majority will vote ``in the wrong way'';
this event is captured by the inequality
$\sum_{i = 1}^{r} X_i \ge r/2$. If $\overline X = \sum_{i = 1}^{r} X_i / r$,
then $A_m = (\overline X \ge 1 / 2)$, while
the expectation satisfies $\Expect \overline X \le 1 / 4$.
We have
\begin{equation*}
\Prob[A_m] =
\Prob[\overline X \ge 1/2] \le
\Prob[\overline X - \Expect \overline X \ge 1/4] \le
\exp\left\{ - 2 \cdot \left(\frac{1}{4}\right)^2 \cdot r \right\} =
e^{-r/8},
\end{equation*}
where the last inequality is the one-sided Chernoff bound
(see, e.g.,~\cite[Section~D.1.2.3]{ChernoffRef}).
Therefore, the probability of Algorithm~\ref{algo:discrete}
giving a bad estimate is upper-bounded by
\begin{equation*}
\Prob[A_1 \cup \ldots \cup A_{\lastm}] \le \lastm \cdot e^{-r/8};
\end{equation*}
to ensure that it is at most~$\alpha$, we pick the smallest $r$ such that
the right-hand side does not exceed~$\alpha$:
\begin{equation}
\label{eq:choice-majority}
r =
\left\lceil
    8 \cdot \ln\left(
            \frac{1}{\alpha} \cdot
            \lfloor n - \log G \rfloor
    \right)
\right\rceil
=
\left\lceil
    8 \cdot \ln\left(
            \frac{1}{\alpha} \cdot
            \lfloor n - 2 \log(\sqrt{a + 1} + 1) \rfloor
    \right)
\right\rceil.
\end{equation}
This completes our analysis.

\end{document}